\documentclass{lmcs} 
\ifdefined\pdfoutput\pdfoutput=1\fi
\usepackage[utf8]{inputenc}

\usepackage{lastpage}
\lmcsdoi{22}{4}{9}
\lmcsheading{}{\pageref{LastPage}}{}{}%
{Apr.~08,~2025}{Oct.~27,~2026}{}


\usepackage{hyperref}
\usepackage{iftex}
\usepackage{amsmath,amssymb,amsthm}
\usepackage{xspace}
\usepackage{booktabs}
\usepackage{float}
\usepackage{placeins}
\usepackage{tikz}
\usetikzlibrary{positioning}
\usetikzlibrary{trees}
\usetikzlibrary{shapes}
\usetikzlibrary{fit}
\usetikzlibrary{shadows}
\usetikzlibrary{backgrounds}
\usetikzlibrary{arrows,automata}

\tikzset{
	n/.style= {circle,fill,inner sep=1.5pt,node distance=2cm}
	,acc/.style={circle,draw,inner sep=3pt,node distance=2cm}
	,phantom/.style={circle},
	,arr/.style={->, >=stealth, semithick, shorten <= 3pt, shorten >= 3pt}
}

\theoremstyle{plain}

\newcommand{\N}{\mathbb{N}}
\newcommand{\Proc}{\mathbb{P}}
\newcommand{\Alp}{\Sigma}
\newcommand{\dom}{\mathit{dom}}
\newcommand{\compdom}{\mathit{Fdom}}
\newcommand{\A}{\mathcal{A}}
\newcommand{\B}{\mathcal{B}}
\newcommand{\Acc}{\mathrm{Acc}}
\newcommand{\Lang}{\mathcal{L}}
\newcommand{\swi}{\mathrm{sc}}
\newcommand{\nextcha}[2]{{#1}{{+}_{_{#2}}}1}
\newcommand{\nextset}{\mathit{inc}}
\newcommand{\pref}{\mathit{pref}}
\newcommand{\toggle}{\mathrm{tg}}
\newcommand{\highl}{\mathrm{hl}}
\newcommand{\RA}{RAA\xspace}
\newcommand{\cost}[2]{\gamma^{#1}(#2)}
\newcommand{\costpas}[2]{\gamma^{#1}_p(#2)}
\newcommand{\costact}[2]{\gamma^{#1}_a(#2)}

\makeatletter

\renewcommand*{\nextcha}[2]{{%
		\mbox{%
			{${#1}$}%
			{{\vbox{\offinterlineskip%
						\hbox{\hskip.9ex
							\scriptsize{${#2}$}}\vskip-.6ex\hbox{\normalsize{${+}$}}}}}%
			{${1}$}%
		}%
}}%

\newcommand*{\prevcha}[2]{{%
		\mbox{%
			{${#1}$}%
			{{\vbox{\offinterlineskip%
						\hbox{\hskip.9ex
							\scriptsize{${#2}$}}\vskip-.6ex\hbox{\normalsize{${-}$}}}}}%
			{${1}$}%
		}%
}}%

\newcommand\maxstar{c^{0^\star}}
\newcommand\minstar{c^{1^\star}}

\begin{document}

		\title{Adding Reconfiguration to Zielonka's Asynchronous Automata}
		\titlecomment{An earlier version appeared in Gandalf 2024.}
		\specialissue{Selected Papers of the 15th International Symposium on Games, Automata, Logics, and Formal Verification (GandALF 2024)}
		\thanks{Supported by the ERC Consolidator grant D-SynMA (No.
		772459), by the Swedish research council (VR) project (No.
		2020-04963) and by the Wallenberg AI, Autonomous Systems and Software 
        Program (WASP) funded by the Knut and Alice Wallenberg Foundation. 
        M.~Lehaut was also partially supported by NCN grant 2021/41/B/ST6/00535.}	

		\author[M.~Lehaut]{Mathieu Lehaut\lmcsorcid{0000-0002-6205-0682}}
		\author[N.~Piterman]{Nir Piterman\lmcsorcid{0000-0002-8242-5357}}

		\address{University of Gothenburg and Chalmers University of
			Technology, Gothenburg, Sweden}





		\begin{abstract}
			\noindent 	
            We study an extension of Zielonka's (fixed)
			asynchronous automata called reconfigurable asynchronous automata
			where processes can dynamically change who they communicate with.
			We show that reconfigurable asynchronous automata are not more
			expressive than fixed asynchronous automata by giving
			translations from one to the other.
			However, going from reconfigurable to fixed comes at the cost of
			disseminating communication (and knowledge) to all processes in
			the system.
			We then show that this is unavoidable by describing a language
			accepted by a reconfigurable automaton such that in every
			equivalent fixed automaton, every process must either be aware of
			all communication or be irrelevant.
		\end{abstract}

		\maketitle

\section{Introduction}
\label{sec:intro}

In recent years, computation devices have become so widely available
that they are now everywhere.
They are lighter, cheaper, prevalent, and, ultimately, mobile.
Sensor networks, multi-agent systems, and robot teams use mobile and
ad-hoc networks.
In such networks, participants/agents/processes come and go and change
the communication configuration based on need, location, and various
restrictions.
These systems force us to consider how communication changes when
participants are numerous, mobile, and required to collaborate.

We consider a canonical formalism in language theory for distributed
systems with a fixed communication structure~--~Zielonka's
\emph{asynchronous automata}.
These are a well known model that supports distribution of language
recognition under a fixed communication topology.
In this model, a number of processes are each connected to some fixed
set of channels.
Those channels enable a symmetric, multiprocess kind of communication called synchronization.
During such a communication, all processes involved share their local
states with each other, and then progress to a new state.
Another feature of this model is that a communication can only happen
if all processes involved are ready for it; if even a single process
does not accept then the communication is blocked.
The model is especially interesting due to Zielonka's seminal result on
the ability to distribute languages in this model \cite{Zielonka87}.
Zielonka's result starts from a given regular language and a target
(fixed) distribution of the alphabet.
He then shows that if the deterministic automaton for the language
satisfies a simple condition saying that independent communications can commute,
then the language can be distributed and accepted by a distributed team of
processes.
Zielonka's quite involved construction has been revisited and optimized
several times, let us cite e.g.
\cite{mukund1997keeping,genest2006constructing,genest2010optimal} for
the general construction and \cite{krishna2013quadratic} for an example
of a simpler construction in a restricted case.
This result leads to several applications notably in synthesis
\cite{mukund2000synthesizing}, further establishing the usefulness of
this model.

The aim of this paper is to extend the power of asynchronous automata
by giving them reconfigurability.
To this end, processes comprising a system are extended with the
ability to dynamically connect and disconnect from channels after a
communication.
As before, a communication can only occur on a channel if all the
processes that are connected to the channel agree on it, and otherwise
the communication is blocked.
This is the exact notion of communication of asynchronous automata,
except
that processes can now connect and disconnect to channels dynamically
during an execution.
In order to allow more than mere synchronization on the channel,
communications are extended by a data value, which corresponds to the
state sharing of asynchronous automata.
We call this variant \emph{reconfigurable asynchronous automata}.
They are inspired by attribute-based communication calculus
\cite{alrahman2015calculus,abd2019calculus} and channeled transition
systems \cite{AlrahmanP21},
though they are much simpler than those and adapted to the context of
asynchronous automata.
To prevent confusions, we sometimes refer to the base variant as
\emph{fixed} (as opposed to reconfigurable) asynchronous automata.

With the definition of this new extension, the first natural question
is whether reconfigurable asynchronous automata are more expressive
than the fixed variant.
To this we answer negatively by showing how to translate from one model
to the other.
Going from fixed to reconfigurable is easy.
We also show that if the fixed asynchronous automaton has local
transitions, i.e. the next state of a process only depends on its own
current state and not on the state of others, then it corresponds to a
reconfigurable asynchronous automaton that does not use data values
during communications.
The other direction is also relatively easy, however with an important
caveat:
every process in the fixed automaton participates in every
communication, autonomously deciding which communications to ignore and
which to act upon.

With the two models being equi-expressive, the second natural question
is what does adding reconfigurability actually bring?
It is well known that non-deterministic finite automata are as
expressive as deterministic ones, but can be exponentially smaller in
size.
In the case of fixed versus reconfigurable asynchronous automata, the
gain is not in size, but in the communication structure.
As explained just before, our translation from reconfigurable to fixed
asynchronous automata results in essentially sharing all information to
all parts of the system, and letting each process decide whether that
information is actually needed.
This is undesirable for many reasons.
First, in real systems, each communication takes time and costs energy
to process, so one should not waste resources sending information that
would be useless to some process.
Second, for privacy reasons, it is obviously not desirable that every
process in the system has access to every communication; we would
rather that a process only receives information based on its
``need-to-know''.
Third, it implies that every process is always connected to every
other process in the system, which is not good in systems with a high
number of participants that only require a small number of connections
at each time.
Fourth, in several cases, algorithms applied to asynchronous automata rely (or use)
sparse communication structure for efficiency or simplicity \cite{krishna2013quadratic,muscholl2014distributed}. 
In one case, we have shown that efficiency is preserved under ``sparse'' reconfiguration \cite{DBLP:conf/atva/HausmannLP24}. This may be lost if we saturate communication to simulate reconfiguration. 

We then show that this sharing is, in general, unavoidable.
We suggest a language that can be recognized by reconfigurable
asynchronous automata but for which any equivalent fixed asynchronous
automaton has the pitfall described above.
In this language, using reconfigurable communication, processes
actively connect and
disconnect from channels and keep themselves informed only about
crucial information.
Throughout, processes are connected to a very small number of channels
that is independent of system size.
However, some (changing) channels are used for coordination of how to
connect and disconnect from the other channels.
We show that for asynchronous automata to recognize the same language,
some
processes must be connected to the full set of channels and be informed
of
everything.
What's more, every process that is not connected to the full set of
channels can be made trivial by accepting unconditionally all possible
communication on channels that they are connected to.
We also show that the system contains a subsystem performing the same
computation in which the processes that are not fully connected are
completely trivial: the system does not need them at all to perform
exactly the same computation.

The rest of the paper is organized as follows.
In Section~\ref{sec:defs} we recall the definition of Zielonka's
asynchronous automata (AA)
and give the definition of reconfigurable asynchronous automata (RAA).
In Section~\ref{sec:translations} we give the translations between the
models and show that the data of reconfigurable asynchronous automata
correspond to the global transitions of fixed asynchronous automata.
We then show in Section~\ref{sec:lower bound} that in every translation
that removes the
reconfigurability, all processes either know everything or are trivial.
Section~\ref{sec:cost} introduces a discussion on the cost of
communication and gives an alternative construction.
Section~\ref{sec:other reconfiguration} discusses some other approaches to reconfiguraiton. 
Finally, we conclude and discuss our results in Section~\ref{sec:conc}.

The paper extends our earlier publication in GandALF 2024
\cite{lehaut2023adding}.
We extend the results by introducing a measure for the cost of
communications, analyzing the cost of communication in the different
models, and introducing an alternative construction of an RAA that
challenges AA, and showing that it changes the balance in the trade-off
between passive and active communication.

\section{Definitions}
\label{sec:defs}

In this section we introduce the two models of interest for this work, asynchronous automata and their reconfigurable variant.

\subsection{Fixed Communication Structure}

We introduce the classic definition of asynchronous automata and give some examples. We distinguish between the \emph{global} and \emph{local} versions differing on the amount of data transferred when communication occurs.  

\subsubsection{Distributed Alphabets}
We fix a finite set $\Proc$ of processes.
Let $\Alp$ be a finite alphabet, and $\dom: \Alp \to 2^\Proc$ a domain
function associating each letter with the subset of processes listening
to that letter.
The pair $(\Alp,\dom)$ is called a distributed alphabet.
We let $\dom^{-1}(p) = \{a \in \Alp \mid p \in \dom(a)\}$.
It induces a binary independence relation $I$ in the following way:
$(a,b) \in I \Leftrightarrow \dom(a) \cap \dom(b) = \emptyset$.
Two words $u = u_1 \dots u_n$ and $v = v_1 \dots v_n$ are said to be
equivalent, denoted by $u \sim v$, if one can start from $u$,
repeatedly switch two consecutive independent letters, and end up with
$v$.
Let us denote by $[u]$ the equivalence class of a word $u$.
Let $\A = (Q,\Alp,q_0,\Delta,F)$ be a deterministic automaton over
$\Alp$.
We say that $\A$ is $I$-diamond if for all pairs of independent letters
$(a,b) \in I$ and all states $q \in Q$, we have $\Delta(q,ab) =
\Delta(q,ba)$.
If $\A$ has this property, then a word $u$ is accepted by $\A$ if and
only if all words in $[u]$ are accepted.
Zielonka's result states that an $I$-diamond automaton can be
distributed to processes who are connected to channels according to
$\dom$ \cite{Zielonka87}.

\subsubsection{Asynchronous Automata}
An \emph{asynchronous automaton} (in short: AA) \cite{Zielonka87} over
distributed alphabet $(\Alp,\dom)$ and processes $\Proc$  is a tuple
$$\B = ((S_p)_{p \in \Proc}, (s^0_p)_{p \in \Proc}, (\delta_a)_{a \in
	\Alp}, \Acc)$$
\noindent
such that:
\begin{itemize}
	\item $S_p$ is the finite set of states for process $p$, and $s^0_p
	\in S_p$ is its initial state,
	\item $\delta_a: \prod_{p \in \dom(a)} S_p \to \prod_{p \in \dom(a)}
	S_p$ is a partial transition function for letter $a$ that only
	depends on the states of processes in $\dom(a)$ and leaves those
	outside unchanged,
	\item $\Acc \subseteq \prod_{p \in \Proc} S_p$ is a set of accepting
	global states, as defined below.
\end{itemize}
A global state of the automaton is of the form $\textbf{s} = (s_p)_{p
\in \Proc}$, giving the state of each process.
For every such global state and every subset $P \subseteq \Proc$, we
denote by $\textbf{s}\downarrow_P = (s_p)_{p \in P}$ the subset of
$\textbf{s}$ of states from processes in $P$.
Then a run of $\B$ is a sequence $\textbf{s}_0 a_1 \textbf{s}_1 a_2
\dots \textbf{s}_n$ where for all $0 < i \leq n$, $\textbf{s}_i \in
\prod_{p \in \Proc} S_p$, $a_i \in \Alp$, satisfying $\textbf{s}_0 =
(s^0_p)_{p \in \Proc}$ and the following relation:
\[
\textbf{s}_{i}\downarrow_{\dom(a_i)} =
\delta_{a_i}(\textbf{s}_{i-1}\downarrow_{\dom(a_i)}) \text{ and }
\textbf{s}_{i}\downarrow_{\Proc\setminus\dom(a_i)} =
\textbf{s}_{i-1}\downarrow_{\Proc\setminus\dom(a_i)}
\]
A run is accepting if $\textbf{s}_n$ belongs to $\Acc$.
The word $a_1 a_2 \dots$ is accepted by $\B$ if such an accepting run
exists (note that automata are deterministic but runs on certain words
may not exist).
The language of $\B$, denoted by $\Lang(\B)$, is the set of words
accepted by $\B$.
For the rest of this paper, we will drop the $\Acc$ component as we
focus more on the runs themselves over whether they can reach a certain
target.
That is,  we assume that $\Acc = \prod_{p \in \Proc} S_p$.
This restricts the languages that can be recognized by asynchronous
automata but still allows us to prove all our results.

\begin{exa}\label{exa:AA}
	We give an example of an asynchronous automaton $\B$ in
	Figure~\ref{example:AA}.
	There are three letters $\Alp = \{a,b,c\}$ distributed over three
	processes $\Proc = \{p,q,r\}$ with the domain:
	$\dom(a) = \{p\}, \dom(b) = \{q\}, \dom(c) = \Proc$.
	An example of a run is the following sequence:
	\[
	(s_1, t_1, u_1)~a~(s_2, t_1, u_1)~b~(s_2, t_2, u_1)~b~(s_2, t_1,
	u_1)~a~(s_1, t_1, u_1)~c~(s_1, t_1, u_1)
	\]
	which gives $abbac$ as a word in $\Lang(\B)$.
	More generally, $\B$ accepts all words where all occurrences of $c$
	are
	such that there are an even number of $a$'s and an even number of
	$b$'s
	in the prefix before the $c$ occurrence. That is,

	$$\Lang(\B) = \left \{ v_0\ldots v_n \in \{a,b,c\}^*  \left |~
	\begin{array}{r}
		\forall i~.~ v_i=c \mbox{ implies } a_\sharp(v_0\ldots v_i) =_{mod
		2} 0
		\\
		\mbox{ and }
		b_\sharp(v_0\ldots v_i) =_{mod 2}0
	\end{array}
	\right .
	\right \},$$
	\noindent
	where $\sigma_\sharp(w)$ is the number of occurrences of letter
	$\sigma$ in word $w$.
\end{exa}

\begin{figure}
	\tikzset{every state/.style={minimum size=15pt}}
	\begin{center}
		\begin{tikzpicture}[
			auto,
			node distance=1.5cm,
			semithick
			]
			\node[state,initial left,initial text=] (a1) {$s_1$};
			\node[draw,rectangle] (p) [above = 0.5cm of a1] {$p$};
			\node[] (lp) [above = 0.5mm of a1] {$a,c$};
			\node[state] (a2) [below of=a1] {$s_2$};
			\node[state,initial left,initial text=] (b1) [right = 2cm of a1]
			{$t_1$};
			\node[draw,rectangle] (q) [above = 0.5cm of b1] {$q$};
			\node[] (lq) [above = 0.0mm of b1] {$b,c$};
			\node[state] (b2) [below of=b1] {$t_2$};
			\node[state,initial left,initial text=] (c1) [right = 2cm of b1]
			{$u_1$};
			\node[draw,rectangle] (r) [above = 0.5cm of c1] {$r$};
			\node[] (lr) [above = 1.3mm of c1] {$c$};

			\path[->] (a1) edge [loop right] node [right] {$c$} (a1);
			\path[->] (a1) edge [bend right] node [left] {$a$} (a2);
			\path[->] (a2) edge [bend right] node [right] {$a$} (a1);
			\path[->] (b1) edge [loop right] node [right] {$c$} (b1);
			\path[->] (b1) edge [bend right] node [left] {$b$} (b2);
			\path[->] (b2) edge [bend right] node [right] {$b$} (b1);
			\path[->] (c1) edge [loop right] node [right] {$c$} (c1);
		\end{tikzpicture}
		\caption{An asynchronous automaton $\B$ over three
		processes $p$, $q$, and $r$. Below each process is the list of letters that contain this process in their domain, and below that is the automaton associated with the process.}\label{example:AA}
	\end{center}
\end{figure}

\subsubsection{Local Asynchronous Automata}
We also define a weaker version of asynchronous automata, called
\emph{local} asynchronous automata (short: LAA or local AA), in which
the transition function is local to each process, and therefore
independent with respect to the states of all other processes.
To avoid confusion, we sometimes refer to normal asynchronous automata
as defined earlier as \emph{global} asynchronous automata (or global
AA), though by default AA refers to global AA.

A \emph{local asynchronous automaton} over $(\Alp,\dom)$ and $\Proc$ is
a tuple $$\B = ((S_p)_{p \in \Proc}, (s^0_p)_{p \in \Proc},
(\delta_p)_{p \in \Proc}),$$
\noindent
where $S_p$ and $s^0_p$ are defined as before, and $\delta_p: S_p
\times \dom^{-1}(p) \to S_p$ is the transition function of process $p$.
A run of $\B$ is a sequence $\textbf{s}_0 a_1 \textbf{s}_1 a_2 \dots
\textbf{s}_n$ where $\textbf{s}_0 = (s^0_p)_{p \in \Proc}$ and for all
$0 < i \leq n$, $\textbf{s}_i = (s_i^p)_{p \in \Proc} \in \prod_{p \in
\Proc} S_p$, $a_i \in \Alp$, satisfying the following relation:
\begin{align*}
	s_{i}^p =
	\begin{cases}
		\delta_{p}(s_{i-1}^p,a_i) &\text{ if $p \in \dom(a_i)$,}\\
		s_{i-1}^p &\text{ otherwise.}
	\end{cases}
\end{align*}
In other words, a run in a local AA is a run on the direct product of the automata of each process.

Observe that a local AA is a syntactic restriction of global AA.
There are languages recognizable by global AA that cannot be recognized
by local AA, because intuitively it would be impossible to make a
process react differently to the same communication based on
differences observed by another process.
For example, take $\Alp = \{a,\bar{a},b,c,\bar{c}\}$ and two processes
$p,q$ such that $p$ listens to $a,\bar{a},b$ and $q$ listens to
$c,\bar{c},b$.
Then take language $L = \{abc,\bar{a}b\bar{c}\}$. One can easily see
that $L$ can be recognized by a global AA but by no local AA.
Similarly, we could  modify the AA in Example~\ref{exa:AA} so that
it accepts the language
$$\Lang(\B) = \left \{ v_0\ldots v_n \in \{a,b,c\}^*  \left |~
\begin{array}{l}
	\forall i~.~ v_i=c \mbox{ implies} \\
	a_\sharp(v_0\ldots v_i)+
	b_\sharp(v_0\ldots v_i) =_{mod 2}0
\end{array}
\right .
\right \}.$$
We do this by replacing the transitions on $c$ as follows. 
We have 
$\delta((s_1,t_1,u_1),c)=(s_1,t_1,u_1)$ and
$\delta((s_2,t_2,u_1),c)=(s_2,t_2,u_2)$.
Namely, allow $c$ transitions only when both $p$ and $q$ agree on the
parity of the number of $a$s and $b$s that they have seen so far.
In particular, Zielonka's distribution result \cite{Zielonka87} no
longer holds for local AA.
Note that the automaton given in Example~\ref{exa:AA} is local.

\subsection{Reconfigurable Communication}

Let us consider here a model where the communication structure is not
fixed, and can be modified dynamically during a run.
Namely, in AA, the alphabet and its distribution are given. This fixes a connection between processes and certain letters.
Contrarily, here, processes come with some initial connectivity, but are aware of the existence of all letters / communication means.
During execution, processes may decide to change the letters they react to, 
either by adding further connections or removing them. 

As before, let us fix a finite set $\Proc$ of processes. 
Let us as well fix a finite set $C$ of channels, with a role similar to
the alphabet $\Alp$ of the previous section.
Here, the function $\dom$ is replaced by a state-dependent listening
function through which processes reconfigure their communication
interfaces depending on their current state.
Finally, let $T$ be a finite set of message
contents.
The intuition behind $T$ is to abstract the state-sharing part of a
communication to allow us to define each process' transition function
independently of other processes.
We emphasize that using message contents to exchange data rather than including, for example, a definition of a transition that is parameterized with the processes that are connected to it, has nothing to do with reconfigurability and is
just a way to simplify definitions.
The reconfigurability comes only from the previously mentionned listening function.

\subsubsection{Reconfigurable Asynchronous Automata}
A \emph{reconfigurable asynchronous automaton} (in short: \RA) over $C$
is a tuple $\A = (S, s^0,
\Delta, L)$ where:
\begin{itemize}
	\item $S$ is a set of states, $s^0 \in S$ being the initial state,
	\item $\Delta: S \times (T \times C) \to S$ is the partial transition
	function, where $\Delta(s,(t,c)) = s'$ means going from state $s$ to
	$s'$ after having a message on channel $c$ with content $t$.
	We write $(s,(t,c),s')\in \Delta$ for $\Delta(s,(t,c))=s'$.
	\item $L: S \to 2^C$ is a listening function such that $c \in L(s)$ if
	there is a transition of the form $(s,(t,c),s') \in \Delta$, i.e.
	state
	$s$ must be listening to channel $c$ if there is some transition from
	$s$ involving a message on $c$.
\end{itemize}
A run of $\A$ is a sequence $s_0 m_1 s_1 m_2 \dots s_n$ starting from
the initial state $s_0 = s^0$ and where for all $0 < i \leq n, m_i \in
T \times C$ and $\Delta(s_{i-1}, m_i )= s_{i}$.
The language of $\A$, denoted by $\Lang(\A)$, is the set of words over
$C$ of the form $c_0 c_1 \dots$ such that there exists a run of the
form $s_0 (t_0, c_0) s_1 (t_1, c_1) \dots$, i.e. we focus only on the
sequence of channels where messages are sent, and drop the states and
message contents.

Intuitively this definition represents the behavior of a single
process, communicating with the outside on channels from $C$.
In order to be able to reconstruct the whole system, we now define the
parallel composition of \RA.

Given a sequence of \RA $(\A_p)_{p \in \Proc}$ with $\A_p = (S_p,
s^0_p, \Delta_p, L_p)$, one can define their parallel composition
$\A_{\parallel \Proc} = (S, s^0, \Delta, L)$:
\begin{itemize}
	\item $S = \prod_{p \in \Proc} S_p$ and $s_0 = (s^0_p)_{p \in \Proc}$,
	\item $L((s_p)_{p \in \Proc}) = \bigcup_{p \in \Proc} L_p(s_p)$,
	\item $\Delta((s_p)_{p \in \Proc}, (t, c)) = (s'_p)_{p \in \Proc}$ if
	the following conditions are met:
	\begin{enumerate}
		\item $\exists p$ s.t. $c \in L_p(s_p)$,
		\item $\forall p$ s.t. $c \in L_p(s_p), (s_p, (t,c), s'_p) \in
		\Delta_p$, and
		\item $\forall p$ s.t. $c \notin L_p(s_p), s'_p = s_p$.
	\end{enumerate}
\end{itemize}
In plain words, there is a transition if all processes listening to the
corresponding channel have a transition with the \emph{same} message
content, with at least one process listening to the channel, whereas
those that do not listen are left unchanged.
Note that if some process listens to that channel but does not
implement the transition, then that transition is blocked.

By convention, an \RA over $C$ and $\Proc$ refers to an \RA of the form
$\A_{\parallel \Proc}$ as described above.
We say that $\A_{\parallel \Proc}$ is \emph{message-deterministic} if, for any global state $s
\in S$ and channel $c \in C$, for all message contents $t,t' \in T$ such that
$\Delta(s,(t,c))$ and $\Delta(s,(t',c))$ are defined, they both lead to the same successor state, 
i.e. $\Delta(s,(t,c)) = \Delta(s,(t',c))$.
That is, from one global state of the system there can not be two different data values that lead to different states.
This restriction becomes important when we consider translation from RAA to AA, as without message determinism, AA would have to support nondeterministic transitions. 
We believe that this is a reasonable restriction as the contents of the message jointly ``arises'' from the states of all processes.
In fact, we would not expect the same global state to support multiple data values.

\begin{exa}\label{exa:RAA}
	Figure~\ref{example:CTS} shows an example of \RA over channels $C =
	\{a,b,c\}$ and three processes $\Proc = \{p,q,r\}$.
	Here we take $T = \{t\}$ as the set of message contents, so for
	readability purposes it is omitted from the transitions.
	Note that when process $p$ is in state $s_2$, it is listening to
	channel $c$ but no $c$-transition is implemented, therefore a
	communication on $c$ is impossible (similarly for $q$ and $t_2$).
	So the only way a communication happens on $c$ is when $p$ and $q$
	are in $s_1$ and $t_1$ respectively, which means only process $r$
	listens to $c$.
	It is then easy to see that this \RA accepts the same language as the
	AA given in Figure~\ref{example:AA}.
	Note that it does so without $p$ or $q$ ever taking part in a
	communication on $c$, contrary to the previous example.
\end{exa}

\begin{figure}
	\tikzset{every state/.style={minimum size=15pt}}
	\begin{center}
		\begin{tikzpicture}[
			auto,
			node distance=1.5cm,
			semithick
			]
			\node[state,initial left,initial text=] (a1) {$s_1$};
			\node[] (La1) [right = 0.2cm of a1] {$a$};
			\node[draw,rectangle] (p) [above = 0.5cm of a1] {$p$};
			\node[state] (a2) [below of=a1] {$s_2$};
			\node[] (La2) [right = 0.2cm of a2] {$a,c$};
			\node[state,initial left,initial text=] (b1) [right = 2cm of a1]
			{$t_1$};
			\node[] (Lb1) [right = 0.2cm of b1] {$b$};
			\node[draw,rectangle] (q) [above = 0.5cm of b1] {$q$};
			\node[state] (b2) [below of=b1] {$t_2$};
			\node[] (Lb2) [right = 0.2cm of b2] {$b,c$};
			\node[state,initial left,initial text=] (c1) [right = 2cm of b1]
			{$u_1$};
			\node[] (Lc1) [right = 0.2cm of c1] {$c$};
			\node[draw,rectangle] (r) [above = 0.5cm of c1] {$r$};

			\path[->] (a1) edge [bend right] node [left] {$a$} (a2);
			\path[->] (a2) edge [bend right] node [right] {$a$} (a1);
			\path[->] (b1) edge [bend right] node [left] {$b$} (b2);
			\path[->] (b2) edge [bend right] node [right] {$b$} (b1);
			\path[->] (c1) edge [loop below] node [right] {$c$} (c1);
		\end{tikzpicture}
		\caption{An RAA $\A$ over three processes. The listening function
		is given to the right of each state.}\label{example:CTS}
	\end{center}
\end{figure}

\section{From Fixed to Reconfigurable and Back}
\label{sec:translations}

We now focus on comparing the expressive power of these two formalisms.
We show translations in both directions.
The translation from fixed to reconfigurable asynchronous automata corresponds to a change of syntax.
The translation from reconfigurable to fixed asynchronous automata introduces a dissemination of the data throughout the system. 
For the rest of this section, we fix a finite set $\Proc$ of processes.

\subsection{Fixed AA to Reconfigurable AA}

Let $(\Alp,\dom)$ be a distributed alphabet, and let $\B$ be an AA over
it.
One can construct an \RA $\A_{\parallel \Proc}$ with $\Alp$ as set of
channels that recognizes the same language as $\B$.

The intuition is as follows.
The listening function of each process is the same for all states: each
process always listens to the channels that have this process in their
domain. 
The only part that is not straightforward to simulate is that a
transition of an AA depends on the states of all processes in the
domain of the corresponding letter.
Therefore each process in the \RA needs to share their state via
message content to all others when simulating a transition.

\begin{thm}
	Every language recognized by an AA over $(\Alp,\dom)$ and $\Proc$ can
	be recognized
	by an \RA with set of channels $\Alp$ and processes $\Proc$.
	\label{lemma:AA to CTS}
\end{thm}

\begin{proof}
	Let $\B = ((S_p)_{p \in \Proc}, (s^0_p)_{p \in \Proc}, (\delta_a)_{a
	\in \Alp})$ be an AA as described earlier.
	For the set of messages, we take $T = \bigcup_{a \in \Alp} (\prod_{p
	\in \dom(a)} S_p)$.

	Then let $\A_p = (S_p,s^0_p,\Delta_p,L_p)$ be an \RA for process $p$
	where:
	\begin{itemize}
		\item $L_p(s) = \{a \in \Alp \mid p \in \dom(a)\}$ for all $s \in
		S_p$,
		\item $\Delta_p (s_p, (t, a)) =
		(\delta_a(t)){\hspace*{-0pt}\downarrow_{\{p\}}}
		\text{ if } s_p = t{\hspace*{-0pt}\downarrow_{\{p\}}}$
	\end{itemize}
	i.e. an $a$-transition is possible if and only if the message $t$ is
	the tuple comprising the current states of all processes in
	$\dom(a)$, and all processes then update their state according to
	$\delta_a$.
	Remark that this means $\A_{\parallel \Proc}$ is message-deterministic.

	By construction, one can show inductively that for each run of $\B$,
	there is a
	corresponding run of $\A_{\parallel \Proc}$ where at each point, the
	state of each process $p$ is the same in both runs.
	It follows that $\Lang(\B) \subseteq \Lang(\A_{\parallel \Proc})$ and
	conversely $\A_{\parallel \Proc}$ can only simulate runs of $\B$,
	showing the reverse inclusion.
\end{proof}

\noindent
Note that the size of the constructed \RA lies almost entirely in the
size of $T$, the message contents set, which is $\prod_{p \in \Proc}
S_p$.

For local AA the translation is even more straightforward, as no
message content is required (i.e. $T$ can be reduced to a singleton).

\begin{cor}
	Every language recognized by an LAA over $(\Sigma,dom)$ can be
	recognized by an \RA with set of channels $\Sigma$ and where $|T|=1$.
	\label{corr:LAA to CTS}
\end{cor}

\begin{proof}
	In the case of LAA, the transition $\delta_p$ does not depend on the
	states of other processes. Let $T=\{t\}$.
	We replace the transition $\Delta_p$ in the proof of
	Theorem~\ref{lemma:AA to CTS} by
	$\Delta_p(s_p,(t,a)) = \delta_p(s_p,a)$.
\end{proof}

\subsection{Reconfigurable to Fixed}

Let us now focus on the reverse direction.
Let $(\A_p)_{p \in \Proc}$ be a sequence of \RA over $\Proc$ with set
of channels $C$, and let $\A$ be their parallel composition.
Our goal is to create an AA with alphabet $C$ that recognizes the same
language.
The question that arises is: what should $\dom$ be defined as for the
distributed alphabet $(C,\dom)$?

The solution is to define it as the complete domain function
$\compdom$: $\compdom(a) = \Proc$ for all channels.
In that case, it is simple to build an AA over $(C,\compdom)$ that
simulates $\A$, as each process can simply stutter when they are not
supposed to listen to a channel.

\begin{thm}
	Every language recognized by a message-deterministic \RA over set of
	channels $C$ and processes $\Proc$ can be
	recognized by an AA over $(C,\compdom)$ and the same set of processes.
\end{thm}

\begin{proof}
	Consider $(\A_p)_{p \in \Proc}$, where
	$\A_p=(S_p,s_p^0,\Delta_p,L_p)$, with
	$\A = (S, s^0, \Delta, L)$ being their parallel composition over
	message contents $T$.
	We build $\B = ((Q_p)_{p \in \Proc}, (q^0_p)_{p \in \Proc},
	(\delta_c)_{c \in C})$ as follows:
	\begin{itemize}
		\item for all $p \in \Proc$, $Q_p = S_p$, and $q^0_p = s^0_p$
		\item For channel $c\in C$ we have $\delta_c$ defined as follows.
		$$\delta_c =
		\left \{
		((q_p)_{p \in \Proc}, (q'_p)_{p \in \Proc}) \left |
		\begin{array}{l}
			\exists p\in \Proc. c\in L_p(q_p) \mbox{ and}\\
			\exists t \in T. \forall p \in \Proc.\\
			\quad\mbox{if }c\in L_p(q_p), (q_{p'}, (t,c), q'_{p'}) \in
			\Delta_{p'} \text{ and}\\
			\quad\mbox{if }
			c \notin L_p(q_p), q_p = q'_p
		\end{array}
		\right .
		\right \}$$
	\end{itemize}

	\noindent
	\looseness=1
	Similarly to Theorem~\ref{lemma:AA to CTS}, the construction makes it
	so that any run of $\A$ has a corresponding run of $\B$ where the
	states are identical for each process, and the same in the other
	direction.
\end{proof}

\noindent
Note that message-determinism is necessary for the transition function to be well defined.
Without that assumption, we would obtain non-deterministic AA, which have not been defined in this paper.
Furthermore, note that having global transitions is necessary to ensure all
processes share the same message content $t$.
However if we assume that $T$ is a singleton, then local transitions
suffice.
Additionally, notice that the construction would still work with a set
$T$ of infinite size, so we could consider RAA where processes
synchronize by agreeing on, say, an integer.

\begin{cor}
	Every language recognized by an \RA over $C$ and $\Proc$, where
	$|T|=1$, can be recognized by an LAA over $(C,\compdom)$ and $\Proc$.
\end{cor}

\noindent
We illustrate this construction in Figure~\ref{example:CTStoAA}.
Note that for this particular example the general construction is not
optimal.
For example, process $p$ is made to listen to $b$ but can never block a
communication on this channel with all states having a self-loop on
reading $b$. Thus, one could safely remove the letter $b$ from the
alphabet of $p$.
By doing similarly on other processes, one can get back the AA from
Example~\ref{exa:AA}.

\begin{figure}
	\tikzset{every state/.style={minimum size=15pt}}
	\begin{center}
		\begin{tikzpicture}[
			auto,
			node distance=1.5cm,
			semithick
			]
			\begin{scope}[shift={(-5.5,0)}]
				\node[state] (a1) {$s_1$};
				\draw[<-] (a1) -- node[above] {} ++(-0.5,0.5);
				\node[] (La1) [right = 0.cm of a1] {$a$};
				\node[draw,rectangle] (p) [above = 0.5cm of a1] {$p$};
				\node[state] (a2) [below of=a1] {$s_2$};
				\node[] (La2) [right = 0.cm of a2] {$a,c$};
				\node[state] (b1) [right = 0.8cm of a1] {$t_1$};
				\draw[<-] (b1) -- node[above] {} ++(-0.5,0.5);
				\node[] (Lb1) [right = 0.cm of b1] {$b$};
				\node[draw,rectangle] (q) [above = 0.5cm of b1] {$q$};
				\node[state] (b2) [below of=b1] {$t_2$};
				\node[] (Lb2) [right = 0.cm of b2] {$b,c$};
				\node[state] (c1) [right = 0.8cm of b1] {$u_1$};
				\draw[<-] (c1) -- node[above] {} ++(-0.5,0.5);
				\node[] (Lc1) [right = 0.cm of c1] {$c$};
				\node[draw,rectangle] (r) [above = 0.5cm of c1] {$r$};

				\path[->] (a1) edge [bend right] node [left] {$a$} (a2);
				\path[->] (a2) edge [bend right] node [right] {$a$} (a1);
				\path[->] (b1) edge [bend right] node [left] {$b$} (b2);
				\path[->] (b2) edge [bend right] node [right] {$b$} (b1);
				\path[->] (c1) edge [loop below] node [right] {$c$} (c1);
			\end{scope}
			\begin{scope}
				\node[state] (a1) {{$s_1$}};
				\draw[<-] (a1) -- node[above] {} ++(-0.5,0.5);
				\node[draw,rectangle] (p) [above = 0.5cm of a1] {$p$};
				\node[] (translation) [below left = 1cm of p] {$\Rightarrow$};
				\node[] (lp) [above = 0.cm of a1] {$a,b,c$};
				\node[state] (a2) [below of=a1] {{$s_2$}};
				\node[state] (b1) [right = 1.5cm of a1] {{$t_1$}};
				\draw[<-] (b1) -- node[above] {} ++(-0.5,0.5);
				\node[draw,rectangle] (q) [above = 0.5cm of b1] {$q$};
				\node[] (lq) [above = 0.cm of b1] {$a,b,c$};
				\node[state] (b2) [below of=b1] {$t_2$};
				\node[state] (c1) [right = 1.5cm of b1] {{$u_1$}};
				\draw[<-] (c1) -- node[above] {} ++(-0.5,0.5);
				\node[draw,rectangle] (r) [above = 0.5cm of c1] {$r$};
				\node[] (lr) [above = 0.cm of c1] {$a,b,c$};

				\path[->] (a1) edge [bend right] node [left] {{$a$}} (a2);
				\path[->] (a2) edge [bend right] node [right] {$a$} (a1);
				\path[->] (b1) edge [bend right] node [left] {$b$} (b2);
				\path[->] (b2) edge [bend right] node [right] {$b$} (b1);
				\path[->] (c1) edge [loop below] node [right] {{$c$}} (c1);
				\path[->] (a1) edge [loop right] node [right] {${b,c}$} (a1);
				\path[->] (a2) edge [loop right] node [right] {${b}$} (a2);
				\path[->] (b1) edge [loop right] node [right] {${a,c}$} (b1);
				\path[->] (b2) edge [loop right] node [right] {${a}$} (b2);
				\path[->] (c1) edge [loop right] node [right] {${a,b}$} (c1);
			\end{scope}
		\end{tikzpicture}
		\caption{On the left, the RAA from Example~\ref{exa:RAA}. On
		the right, its translation to an AA.}\label{example:CTStoAA}
	\end{center}
\end{figure}

\noindent
There is an alternative construction that does not require \emph{all}
processes to listen to \emph{all} channels.
If one process does while also storing the state information of every
other processes, then it can simulate the original automaton by itself;
meanwhile every other process can listen to an arbitrary set of
channels as long as they accept every communication.
In other words, one process serves as a centralized executor of the
simulation, while others simply need to be non-blocking.
With a centralized executor there is no point in having computation
done anywhere but in the centralized executor.
By abuse of definition we still refer to such a domain function as a
complete domain.

In the next section we show that there is no hope of finding a
transformation that does not require a complete domain.

\section{Trivializable, Fully Listening, and Trivial}
\label{sec:lower bound}
The method described above is a general method to transform a
reconfigurable asynchronous automaton into an equivalent fixed
asynchronous automaton, with the cost of needing a complete domain
function.
It is of course possible that for some particular examples such a heavy
construction is not needed, and a translation with a much smaller
domain could be possible.
However, we show that there is no better (in terms of channel domain)
\emph{general} translation by giving an example of an \RA
such that every equivalent AA relies on a complete domain.

The idea is to allow every possible subset of channels to be either
fully independent, that is every one of those channels can be used in
parallel, or make them sequentially dependent, that is they can only be
used in a certain order.
This status can be switched by a communication on a separate channel
(that all processes listen to), called the switching channel.
Moreover, after enough switches, a different channel will serve as the
switching channel.
That way, all channels have the opportunity to serve as the switching
channel, given enough switches.
Our construction does not use message contents during communications.
Thus, already the weakest form of \RA ($T$ is a singleton) is enough
for this example.

Technically, this is implemented using the following notions:
\begin{itemize}
    \item 
    At every given moment of time there is a set of channels on which communication is unrestricted and (the complement) set of channels on which communication happens only in a fixed (cyclic) order. 
    Each process stores their current view of what is this set as part of their state space. 
    We store the restricted set rather than (its complement) the unrestricted set.
    The coordination between all processes ensures that all processes have a common view of what is this set at all times. 
    \item 
    At every given moment of time, every process is in charge of one channel. This process ``initiates'' communication on this channel. If this channel needs to respect a given order, this process waits for the communication that enables this channel and only then initiates. 
    Thus, the state space of each process memorizes the channel they are in charge of.
    They need to memorize whether they are awaiting communication on another channel in order to initiate communication on their channel.
    In order to do the latter, we add to the state space of a process the channel on which they await the next communication. 
    \item 
    At every given moment of time, one of the channels (switching channel) serves as a global coordination channel.
    All processes need to maintain the identity of this channel in their state space.
    A communication on this channel changes the set of ordered channels according to some agreed upon order between all possible sets. 
    Once we have gone through all possible sets of ordered channels, the switching channel itself changes.
    The coordination between all processes ensures that all processes have a common view of which channel is the switching channel. 
\end{itemize}

\subsection{Description of the switching RAA}

Let $\Proc = \{p_1, \dots, p_n\}$.
We fix the set of channels $C = \{c_1, \dots, c_n, c_{n+1}\}$, that is, we have one channel
per process and one additional channel to be used as switching channel
(dynamically).

For all $\swi \in  C$ ($\swi$ stands for \emph{switching channel}), fix
$<_\swi$ an arbitrary total order over $2^{C \setminus \{\swi\}}$, with
the only requirement that $\emptyset$ be the minimal element.
Intuitively, a set in $2^{C \setminus \{\swi\}}$ will represent the set
of dependent channels, and a switch will go to the next one with
respect to $<_\swi$.
Let us denote by $\nextset_{<_\swi}: 2^{C \setminus \{\swi\}} \to 2^{C
\setminus \{\swi\}} \cup \{\bot\}$ the function that returns the next
set according
to $<_\swi$ or $\bot$ for the maximal element.

Additionally, for every subset $D \subseteq C$, we fix $\nextcha{}{D}:
C \to C$ a function that cycles through all elements of $D$ and is the
identity on $C\setminus D$.
For convenience we write $\nextcha{d}{D}$ for $\nextcha{}{D}(d)$.
We also define $\prevcha{}{D}:D \to D$ the inverse function and use the
same notation.
Namely, for every $d\in D$ we have $\nextcha{(\prevcha{d}{D})}{D}=d$
and $\prevcha{({\nextcha{d}{D}})}{D}=d$.
We denote by $c_D \in D$ an arbitrary element of $D$.

Finally, we set $T = \{t\}$, and omit the message content
component in transitions.

We build $\A_p = (S_p, s^0_p, \Delta_p, L_p)$ for $p = p_k$ as follows:
\begin{itemize}
	\item $S_p = \{(c, \swi, D, d) \mid c, \swi \in C, D \subseteq C
	\setminus \{\swi\}, d \in D \cup \{c\}\}$, and $s_p^0 = (c_k, c_{n+1},
	\emptyset, c_k)$.

	The first component is the channel assigned to this process,
	initially $c_k$ for process $p_k$, but may change if $c_k$ becomes
	the switching channel.
	The second component is the current switching channel, initialized to
	$c_{n+1}$ for all processes.
	Component $D$ represents the set of channels that are currently
	dependent, and $d$ is the next channel that $\A_k$ is listening to on
	which it is expecting communication.
	Note that if $D$ is the empty set or any singleton, it means there
	are no constraints on communication for every non-switching channel.
	Thus, the behavior of $\A_p$ is the same for all sets $D$ such that
	$|D|\leq 1$.
	\item
	All processes listen to the switching channel and their assigned
	channel, plus the previous one if $D$ contains the assigned channel:

	$$L_p(c, \swi, D, d) =
	\left \{ \begin{array}{l l}
		\{\swi, c , \prevcha{c}{D}\} & \mbox{if }c \in D\\
		\{\swi,c\} & \mbox{if }c \notin D
	\end{array} \right.$$

	\item The transition $\Delta_p$ is the union of the following six sets.
	\begin{align}
		&\{((c, \swi, D, c), c, (c, \swi, D, \prevcha{c}{D}))\}\\
		&\{((c, \swi, D, \prevcha{c}{D}), \prevcha{c}{D}, (c, \swi, D, c))\}
	\end{align}
	The first two kinds of transitions handle the independence of all
	channels in $C\setminus D$ and the cycling through the channels of
	$D$.
	If $c\notin D$ then $c=\prevcha{c}{D}$. In this case, the first two
	sets simply say that a transition on $c$ is always possible.
	If $c\in D$, then the process awaits until it gets a message on
	$\prevcha{c}{D}$ and then is ready to interact on $c$.
	After interaction on $c$ it awaits another interaction on
	$\prevcha{c}{D}$.
	It follows that all the processes owning the channels in $D$ enforce
	together the cyclic order on the messages in
	$D$.
	This part is further illustrated in
	Figure~\ref{fig:order
		in d}.

	Remaining transitions describe what happens when a switch occurs.
	\begin{align}
		&\{ ((c, \swi, D, d), \swi, (c, \swi, D', c)) ~|~ D' =
		\nextset_{<_\swi}(D) \neq \bot \mbox{ and } c=c_{D'} \}\\
		&\{ ((c, \swi, D, d), \swi, (c, \swi, D', \prevcha{c}{D'})) ~|~ D'
		= \nextset_{<_\swi}(D) \neq \bot \mbox{ and } c\neq c_{D'} \}
	\end{align}
	Sets three and four describe what happens when the next set according
	to $<_\swi$ is defined.
	In this case, the next set becomes the new set of dependent channels
	$D$.
	Set three handles the case of the process that is in charge of the
	channel becoming the first channel to communicate on the new set
	$\nextset_{<_\swi}(D)$.
	This process is ready for communication on this first channel.
	The fourth set handles the case of all other processes.
	All other processes are either in charge of channels in $D'$, in which
	case they set themselves to await a communication on the previous in
	$D'$ or they are in charge of channels not in $D'$ in which case, $c$
	and $\prevcha{c}{D'}=c$, and the process is ready to communicate on
	$c$.
	\begin{align}
		&\{((c, \swi, D, d), \swi, (c, \nextcha{\swi}{C}, \emptyset,c)) ~|~
		\nextset_{<_\swi}(D)=\bot \mbox{ and } c\neq \nextcha{\swi}{C}\}\label{equation:next set first}\\
		&\{((c, \swi, D, d), \swi, (\swi,
		\nextcha{\swi}{C},\emptyset,\swi)) ~|~
		\nextset_{<_\swi}(D) = \bot \mbox{ and } c = \nextcha{\swi}{C} \}\label{equation:next set second}
	\end{align}
	Finally, sets five and six describe what happens when the next set
	according to $<_\swi$ is undefined.
	In this case, the next dependent set becomes $\emptyset$.
	Most processes just set the dependent set to $\emptyset$ and allow
	communication on ``their'' channel (\ref{equation:next set first}).
	The process that was in charge of the new switching channel
	$\nextcha{\swi}{C}$ takes over the old switching channel $\swi$ and is
	ready to communicate on it (\ref{equation:next set second}).
	Notice that communications on the switching channel affect all
	processes.
	The change in $D$ and the change of the switching channel is further
	illustrated in Figure~\ref{fig:switching}.
\end{itemize}

\begin{figure}[H]
	\begin{center}
		\includegraphics[width=0.5\linewidth]{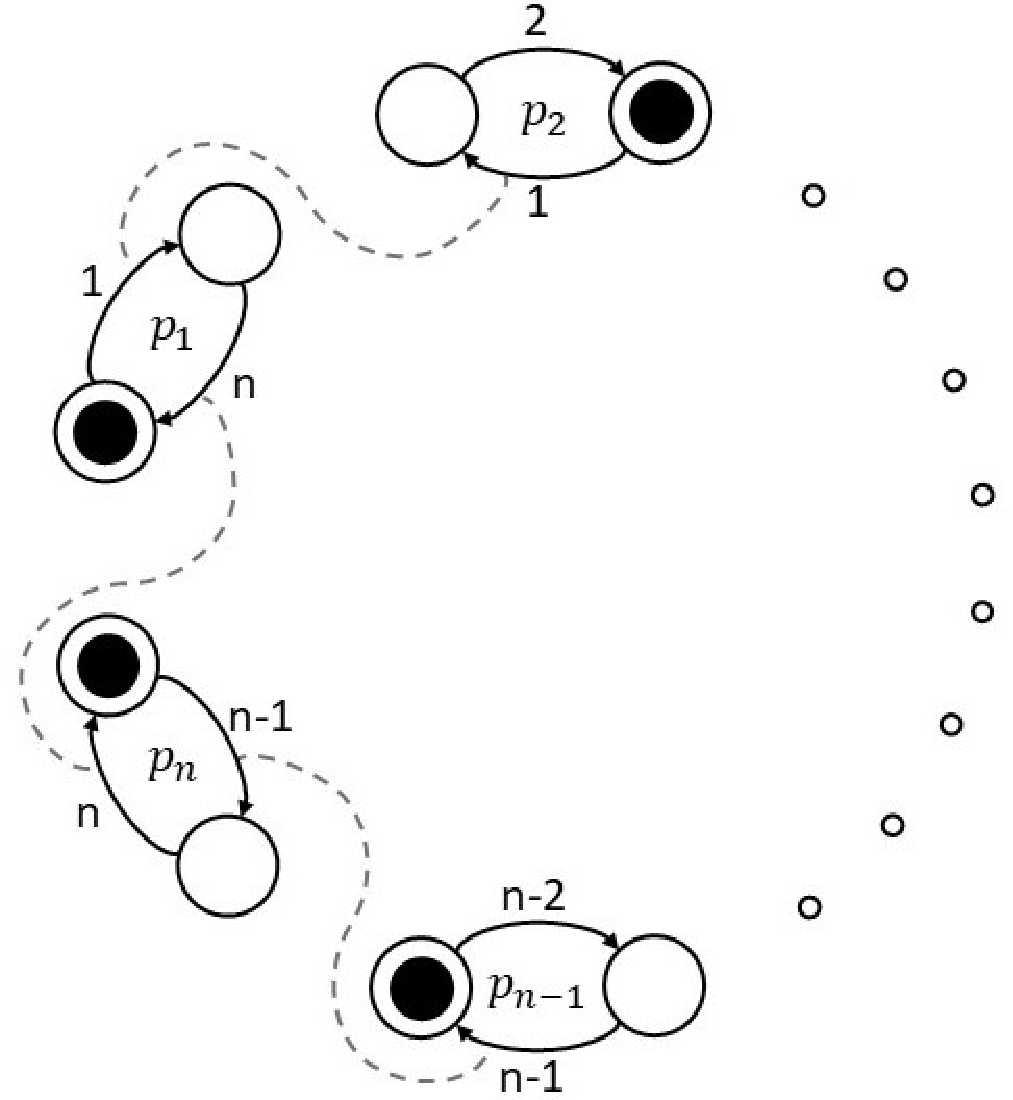}
	\end{center}
	\caption{\label{fig:order in d}Illustration of how the order on the
		channels in $D$ is maintained.
		We consider the case where
		$D=\{1,\ldots,n\}$ and $p_i$ is in charge
		of channel $i$.
		The order between the channels is the
		natural order on $\{1,\ldots, n\}$.
		The black token indicates the current state
		for each process.
		Transitions that are on the same channel
		are connected with a dashed line.
		The system is set up for next communication
		on channel $1$ and all other channels are
		blocked.
		Indeed, both processes listening to channel $1$
		are ready to interact on $1$ ($p_1$ in
		state $(1,n+1,D,1)$ and $p_2$ in state
		$(2,n+1,D,1)$) and for every
		other channel $i>2$ process $i$ is awaiting
		communication on $i-1$ ($p_i$ in state
		$(i,n+1,D,i-1)$) so channel $i$ is
		not enabled.
	}
\end{figure}

\begingroup
\setlength{\intextsep}{14pt plus 0pt minus 0pt}
\begin{figure}[H]
	\begin{center}
		\includegraphics[width=\linewidth]{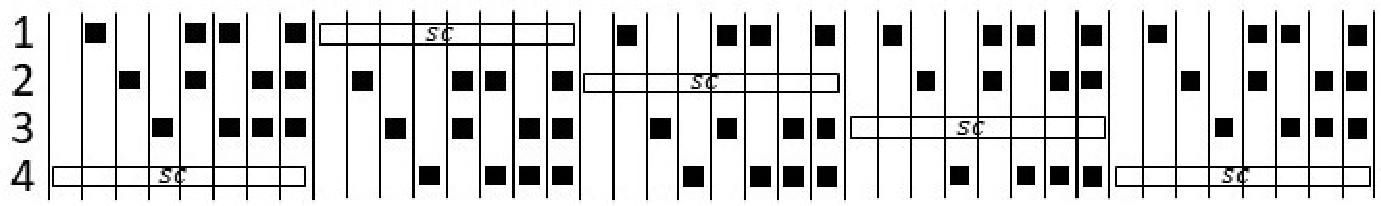}
	\end{center}
	\caption{\label{fig:switching}
		Illustration of how the set $D$ and the switching channel $\swi$
		change whenever there is a communication on $\swi$.
		We consider the case where there are three processes and four
		channels.
		Each column corresponds to the status after one more communication
		on $\swi$.
		Each channel is in turn the switching channel starting with $4$.
		The channels in $D$ at a certain time/column are marked with a
		black box.
		We cycle through the states in $2^{\{1,\ldots, 4\} \setminus \{\swi\}}$
		according to set size first and then lexicographically on the
		sorted set.
	}
\end{figure}
\endgroup

\noindent
An illustration of the whole construction for $n = 2$ (i.e. 2 processes
and 3 channels) is given in Figure~\ref{example:construction}.
There we have processes $\Proc = \{p,q\}$ and channels $C = \{1,2,3\}$.
Initially $p$ is assigned channel $1$ and $q$ channel $2$, while
channel $3$ is the switching channel.
We chose as order for the dependent sets the following order:
$\emptyset <_3 \{1\} <_3 \{2\} <_3 \{1,2\}$.
The blue states illustrate the moment when the dependent set has two
channels that must be used in the correct order ($1 \rightarrow 2
\rightarrow 1 \rightarrow ...$).
The red transitions lead to a change in the switching channel.
At that point, $1$ becomes the new switching channel.
Thus $p$ gets a new assigned channel ($3$, i.e. the previous switching
channel), while $q$ keeps its old assigned channel.
After enough changes of the switching channel, the state cycles back to
the initial state for both.

\begin{figure}[t]
	\tikzset{every state/.style={minimum size=15pt}}
	\begin{center}
		\begin{tikzpicture}[
			auto,
			node distance=1.5cm,
			semithick
			]
			\begin{scope}
				\node[draw,rounded corners,initial left,initial text=] (p1)
				{$1, 3, \emptyset, 1$};
				\node[] (Lp1) [below = 0.cm of p1] {$1,3$};
				\node[draw,rounded corners] (p2) [above right = 0.5cm of p1]
				{$1, 3, \{1\}, 1$};
				\node[] (Lp2) [below = 0.cm of p2] {$1, 3$};
				\node[draw,rounded corners] (p3) [right = 0.5cm of p2] {$1, 3,
				\{2\}, 1$};
				\node[draw,rectangle] (p) [above = 1cm of p3] {$p$};
				\node[] (Lp3) [below = 0.cm of p3] {$1, 3$};
				\node[draw,rounded corners,color=blue] (p4) [below right =
				0.5cm of p3] {$1, 3, \{1,2\}, 1$};
				\node[] (Lp4) [above = 0.cm of p4] {$1, 2, 3$};
				\node[draw,rounded corners,color=blue] (p5) [right = 0.5cm of
				p4] {$1, 3, \{1,2\}, 2$};
				\node[] (Lp5) [above = 0.cm of p5] {$1, 2, 3$};
				\node[draw,rounded corners,color=red] (p6) [below = 1cm of p5]
				{$3, 1, \emptyset, 3$};
				\node[] (Lp6) [below = 0.cm of p6] {$1, 3$};
				\node[] (pdots1) [left of =p6] {$...$};
				\node[] (pdots2) [below right = 1.5cm of p1] {$...$};

				\path[->] (p1) edge [loop above] node [above] {$1$} (p1);
				\path[->] (p1) edge [] node [above left] {$3$} (p2);
				\path[->] (p2) edge [loop above] node [above] {$1$} (p2);
				\path[->] (p2) edge [] node [above] {$3$} (p3);
				\path[->] (p3) edge [loop above] node [above] {$1$} (p3);
				\path[->] (p3) edge [] node [above right] {$3$} (p4);
				\path[->,color=blue] (p4) edge [bend right] node [above] {$1$}
				(p5);
				\path[->,color=blue] (p5) edge [bend right] node [above] {$2$}
				(p4);
				\path[->,color=red] (p4) edge [] node [below left] {$3$} (p6);
				\path[->,color=red] (p5) edge [] node [below right] {$3$} (p6);
				\path[->] (p6) edge [loop right] node [right] {$3$} (p6);
				\path[->] (p6) edge [] node [above] {$1$} (pdots1);

				\draw[dashed, rounded corners] (-1,2.2) rectangle (11,-0.7) {};
				\draw[] (1,2.4) node {Switching channel = $3$};
				\draw[dashed, rounded corners] (-1,-0.8) rectangle (5,-2.4) {};
				\draw[] (1,-2.6) node {Switching channel = $2$};
				\draw[dashed, rounded corners] (5.2,-0.8) rectangle (11,-2.4)
				{};
				\draw[] (7,-2.6) node {Switching channel = $1$};
				\draw[->,color=red] (5.5,-1.5) -- (4.7,-1.5);
				\draw[->,color=red] (2,-1) -- (2,-0.5);
			\end{scope}
			\begin{scope}[shift={(0,-6)}]
				\node[draw,rounded corners,initial left,initial text=] (p1)
				{$2, 3, \emptyset, 2$};
				\node[] (Lp1) [below = 0.cm of p1] {$2,3$};
				\node[draw,rounded corners] (p2) [above right = 0.5cm of p1]
				{$2, 3, \{1\}, 2$};
				\node[] (Lp2) [below = 0.cm of p2] {$2, 3$};
				\node[draw,rounded corners] (p3) [right = 0.5cm of p2] {$2, 3,
				\{2\}, 2$};
				\node[draw,rectangle] (p) [above = 1cm of p3] {$q$};
				\node[] (Lp3) [below = 0.cm of p3] {$2, 3$};
				\node[draw,rounded corners,color=blue] (p4) [below right =
				0.5cm of p3] {$2, 3, \{1,2\}, 1$};
				\node[] (Lp4) [above = 0.cm of p4] {$1, 2, 3$};
				\node[draw,rounded corners,color=blue] (p5) [right = 0.5cm of
				p4] {$2, 3, \{1,2\}, 2$};
				\node[] (Lp5) [above = 0.cm of p5] {$1, 2, 3$};
				\node[draw,rounded corners,color=red] (p6) [below = 1cm of p5]
				{$2, 1, \emptyset, 2$};
				\node[] (Lp6) [below = 0.cm of p6] {$1, 2$};
				\node[] (pdots1) [left of =p6] {$...$};
				\node[] (pdots2) [below right = 1.5cm of p1] {$...$};

				\path[->] (p1) edge [loop above] node [above] {$2$} (p1);
				\path[->] (p1) edge [] node [above left] {$3$} (p2);
				\path[->] (p2) edge [loop above] node [above] {$2$} (p2);
				\path[->] (p2) edge [] node [above] {$3$} (p3);
				\path[->] (p3) edge [loop above] node [above] {$2$} (p3);
				\path[->] (p3) edge [] node [above right] {$3$} (p4);
				\path[->,color=blue] (p4) edge [bend right] node [above] {$1$}
				(p5);
				\path[->,color=blue] (p5) edge [bend right] node [above] {$2$}
				(p4);
				\path[->,color=red] (p4) edge [] node [below left] {$3$} (p6);
				\path[->,color=red] (p5) edge [] node [below right] {$3$} (p6);
				\path[->] (p6) edge [loop right] node [right] {$2$} (p6);
				\path[->] (p6) edge [] node [above] {$1$} (pdots1);

				\draw[dashed, rounded corners] (-1,2.2) rectangle (11,-0.7) {};
				\draw[] (1,2.4) node {Switching channel = $3$};
				\draw[dashed, rounded corners] (-1,-0.8) rectangle (5,-2.4) {};
				\draw[] (1,-2.6) node {Switching channel = $2$};
				\draw[dashed, rounded corners] (5.2,-0.8) rectangle (11,-2.4)
				{};
				\draw[] (7,-2.6) node {Switching channel = $1$};
				\draw[->,color=red] (5.5,-1.5) -- (4.7,-1.5);
				\draw[->,color=red] (2,-1) -- (2,-0.5);
			\end{scope}
		\end{tikzpicture}
		\caption{Illustration of the switching RAA for $n =
		2$.}\label{example:construction}
	\end{center}
\end{figure}

Let $\A$ be the parallel composition of $(\A_p)$.
A state for one process keeps track of 3 channels and one set of
channels.
However, the channel on which they ``await'' communication is either equivalent to their main channel or is the previous channel in the order. Thus a single bit is required to memorize this information.
It follows that the size of a single process is in $O(n^2.2^n)$.
The set of channels is maintained uniform across all processes and so is the identity of the switching channel. 
Furthermore, once we know which is the switching channel and we know which is the process in charge of the least other channel (i.e., channel $1$ or in case that the switching channel is $1$, channel $2$), the rest of the association between the channels and processes respects a cyclic order.
Therefore, the size of the state space of $\A$ is in $O(n^{3}.2^n)$ (rather than $(n^{2}.2^n)^{O(n)})$.

The language $\Lang(\A)$ is complex, but can be decomposed in smaller parts.
First, let $\Lang_D^\swi$ be the language over $C \setminus \{\swi\}$ respecting $D$:
$w \in \Lang_D^\swi$ if the first occurrence of a letter in $D$ in $w$ is $c_D$, 
and for all $d \in D$ that occur in $w$, the next occurrence of $D$ in $w$ (if there is one) is $\nextcha{d}{D}$.
Note the absence of constraints for letters not in $D$.
Second, let $\Lang^\swi = \Lang_{D_0}^\swi \cdot \swi \cdot \Lang_{D_1}^\swi \cdot \swi \cdot \ldots \cdot \swi \cdot \Lang_{D_{2^n-1}}^\swi$, with $D_0 = \emptyset$ and $D_{i+1} = \nextset_{<_\swi}(D_i)$ for all $0 \leq i < 2^n-1$.
This is the language of all words when $\swi$ is the current switching channel.
Finally, we have $\Lang(\A) = (\Lang^{c_{n+1}} \cdot c_{n+1} \cdot \Lang^{c_1} \cdot c_1 \cdot \ldots \cdot \Lang^{c_n} \cdot c_n)^\ast$:
we start with the language where $c_{n+1}$ is the switching channel, then after the last switch $c_1$ assumes this role, and so on until the last switch using $c_n$ at which point we end up back at the beginning.

\FloatBarrier
\subsection{Asynchronous Automata Construction}

We show that an AA that recognizes the same language as $\A$ has the
following property: for each process $p$, either $p$ listens to every
channel ($\dom^{-1}(p) = C$), or from every reachable state there is a
path to a bottom strongly connected component that is complete w.r.t.
$\dom^{-1}(p)$.
That is, for every state $s$ in this bottom SCC and for every channel
in $\dom^{-1}(p)$ the transition $\delta(s,c)$ is defined.
In the former case, we call $p$ \emph{fully-listening}.
In the latter case, we say that $p$ is \emph{trivializable}, as once it
is in this bottom SCC it always includes transitions for all the
channels it listens to.
Thus, $p$ becomes irrelevant to the rest of the computation.

\begin{thm}\label{lemma:main}
	Let $\B$ be an AA such that $\Lang(\B) = \Lang(\A)$. Each process in
	$\B$ is either fully-listening or trivializable.
\end{thm}

\begin{proof}
	Let $\B = ((S_p)_{p \in \Proc}, (s^0_p)_{p \in \Proc}, (\delta_a)_{a
	\in \Alp})$, and let $p \in \Proc$.
	Assume that $p$ is not fully-listening, so let $C_p = \dom^{-1}(p)
	\subsetneq C$.
	In particular, let $c \in C \setminus C_p$ be a channel that $p$ does
	not listen to.

	Let $s_p$ be a reachable state for $p$ in $\B$.
	Then there is $w$ a computation in $\Lang(\B)$ such that $p$ reaches
	state $s_p$ after $w$.
	Consider the same computation in $\A$, and let $\swi$ be the current
	switching channel in $\A$ at the end of $w$.
	Let $\swi \cdot c_1 \cdot ... \cdot c_{n-1} \cdot c$ be the sequence
	of
	channels from $\swi$ to $c$ according to the order from
	$\nextcha{}{C}$.
	Then there is a continuation $w'$ of $w$ of the form $\swi^{k_0}
	\cdot c_1^{k_1} \cdot ... \cdot c_{n-1}^{k_{n-1}}$ with $k_0, \dots,
	k_{n-1} \in \N$ such that:
	\begin{itemize}
		\item $ww' \in \Lang(\A)$,
		\item after $w'$, $c$ is the current switching channel and the
		dependent set $D$ is $\emptyset$.
	\end{itemize}
	From this, every continuation $w'' \in (C \setminus \{c\})^\ast$ is
	still in $\Lang(\A)$ and does not change the switching channel or the
	dependent set.
	In particular, every $w'' \in (\dom^{-1}(p))^*$ maintains that $w\cdot
	w'
	\cdot w''$ is also in $\Lang(\A) =
	\Lang(\B)$.
	Therefore, from the state reached in $\B$ after $ww'$, there is a path
	to a strongly connected component that will implement all transitions
	in $\dom^{-1}(p)$, i.e. a complete one.
\end{proof}

\noindent
A process that is trivializable may become irrelevant.
This means that there are pathological runs where only
fully listening processes are active in the computation while
others passively accept everything.
However, trivializable processes \emph{may} still initially participate
in
the computation.
Nevertheless, for the languages given in this section, we
show that there exists an alternative initial configuration of the
system where trivializable processes actually start
trivialized.
This means that, in essence, all the machinery required for doing the
entire computation is present within the remaining fully listening
processes only.

Given a language $\Lang$ and a word $w$ let $w\backslash \Lang= \{w'
~|~ ww'\in \Lang\}$ and let $\pref(\Lang)=\{ w ~|~ \exists w'~.~ ww'\in
\Lang\}$.
A language $\Lang$ is \emph{repetitive} if for every word $w\in
\pref(\Lang)$
there exists a word $w'$ such that $ww'\backslash \Lang=\Lang$.

\begin{lem}
	The language $\Lang(\A)$ is repetitive.
	\label{lemma:repetitive}
\end{lem}

\begin{proof}
	Consider a word $w$ and the configuration of $\A$ reachable after
	reading $w$.
	All processes in $\A$ agree on the set $D$ and the channel $\swi$.
	Every communication on $\swi$ increases the set of dependent channels
	in the order $<_{\swi}$ until reaching the set $D'$ such that
	$\nextset_{<_\swi}(D')=\bot$.
	An additional communication on $\swi$ then leads to the switching
	channel being updated to $\nextcha{\swi}{C}$.

	So after at most $2^n$ communications on $\swi$ the switching
	channel becomes $\nextcha{\swi}{C}$.
	Let $w_0$ be the word that leads to the switching channel changing.

	For every channel, $c_i$, when the dependent set is $\emptyset$ the
	sequence $(c_i)^{2^n}$ leads to the change of the switching channel
	from $c_i$ to $\nextcha{c_i}{C}$.

	Let $\swi=c^0$, $c^1$, $\ldots$, $c^k$ be the sequence of switching
	channels ending $\nextcha{c^k}{C}=c_{n+1}$.

	It follows that $w_0\cdot (c^1)^{2^n}\cdots (c^k)^{2^n}$ leads $\A$
	to setting the switching channel to $c_{n+1}$.
	At that point all processes in $\A$ are in their initial
	states except for their assigned channel, which is shifted by one.
	Namely, process $p_k$ ends up in state
	$(\prevcha{c_k}{(C \setminus \{c_{n-1}\})},c_{n+1},\emptyset,c_k)$.

	Now let $w_\mathrm{loop} = c_{n+1}^{2^n} \cdot c_1^{2^n} \cdots
	c_n^{2^n}$.
	Each application of $w_\mathrm{loop}$ again shifts assigned channels
	by one.
	So after $n-1$ applications, each process finishes in its initial
	state.
	From this configuration the residue language is $\Lang(\A)$.
\end{proof}

\noindent
Using  repetitiveness we can strengthen our result as follows.
A process is \emph{trivial} if its initial state lies in a bottom
strongly connected component that is complete w.r.t. $\dom^{-1}(p)$.
Given an AA $\B = ((S_p)_{p \in \Proc}, (s^0_p)_{p \in \Proc},
(\delta_a)_{a \in	\Alp})$ and an alternative initial
configuration $\vec{t}=(t^0_p)_{p\in \Proc}$ we denote by $\B(\vec{t})$
the AA $\B(\vec{t}) =  ((S_p)_{p \in \Proc}, (t^0_p)_{p \in \Proc},
(\delta_a)_{a
	\in	\Alp})$.

\begin{thm}
	Let $\B$ be an AA such that $\Lang(\B) = \Lang(\A)$.
	There exists an alternative initial configuration
	$\vec{t}=(t_0)_{p\in\Proc}$ such that $\Lang(\B(\vec{t}))=\Lang(\A)$
	and each process in $\B(\vec{t})$ is either fully-listening or
	trivial.
	\label{lemma:trivial}
\end{thm}

\begin{proof}
	Let $\B$ be an AA equivalent to $\A$.
	By Theorem~\ref{lemma:main} there exists a word $w$ such that after
	reading $w$ all processes in $\B$ that are not fully listening
	reached a bottom SCC, where they accept all communications on all
	channels they are listening to.
	By Lemma~\ref{lemma:repetitive}, there exists a word $w'$ such that
	$ww'\backslash \Lang(\B)=\Lang(\B)$.
	Let $\vec{t}=(t^0_p)_{p\in \Proc}$ be the states that processes in
	$\B$ reach after reading $ww'$.
	Then $\Lang(\B(\vec{t}))=\Lang(\B)$.
	The theorem follows.
\end{proof}

\section{The Cost of Communication}
\label{sec:cost}
In the previous section, we have shown that although they are
equivalent in terms of language expressivity, reconfigurable automata
allow for a much lower communication footprint than fixed automata.
In this section, we introduce a measure that allows us to further study
the differences.
Our measure computes the cost of communication over an execution.
We then use this measure to compare the costs of different executions
in the two models clarifying what is the price of full dissemination
and when (just in terms of communication cost) it becomes worthwhile to
support reconfigurability.
Finally, we give an alternative construction of an RAA (requiring full
distribution) that suggests a different balance of costs than the one
from Section~\ref{sec:lower bound}.

\subsection{A Measure for Communication Cost}
\subsubsection{Measure Definition}
First, let us define a measure that reflects how expensive it is to
maintain channels and communicate on them.
We distinguish two different costs, \emph{passive} and \emph{active}.
The passive cost is defined as the number channels a process is
connected to at the same time; the fewer there are the lower the cost.
For a fixed automaton, this cost is constant over the whole execution.
For a reconfigurable automaton, this cost is computed at each action
and summed up over the execution.
The second kind is the active cost, which refers to the cost incurred
by a communication.
The cost of a communication is defined as the number of participants in
it.

Formally, let us fix some set of channels $C$ and set of processes
$\Proc$,
an AA $\B = ((S_p)_{p \in \Proc}, (s^0_p)_{p \in \Proc}, (\delta_a)_{a
\in \Alp})$ over $\dom$,
and an RAA $\A = ((S_p)_{p \in \Proc}, (s^0_p)_{p \in
\Proc},(\Delta_p)_{p \in \Proc}, (L_p)_{p \in \Proc})$.
Given an execution $w = c_0 c_1 \dots \in C^\ast$, we define the
passive costs $\costpas{\A}{w}$, $\costpas{\B}{w}$ and the active costs
$\costact{\A}{w}$, $\costact{\B}{w}$ as follows:
\begin{align*}
	\costpas{\A}{w} &= \sum_{0 \leq i < |w|} \sum_{p \in \Proc}
	|dom^{-1}(p)| = |w| \cdot \sum_{c \in C} |\dom(c)|\\
	\costpas{\B}{w} &= \sum_{0 \leq i < |w|} \sum_{p \in \Proc}
	|L_p(s_p^i)|\\
	\costact{\A}{w} &= \sum_{0 \leq i < |w|} |\dom(c_i)|\\
	\costact{\A}{w} &= \sum_{0 \leq i < |w|} |\{p \mid c_i \in
	L_p(s_p^i)\}|
\end{align*}
where $s_p^i$ denotes the state reached by process $p$ in $\A$ after
the prefix $c_0 \dots c_{i-1}$ of $w$.
We define the costs $\cost{\A}{w}$ and $\cost{\B}{w}$ as their sums:
\begin{align*}
	\cost{\A}{w} &= \costpas{\A}{w} + \costact{\A}{w}\\
	\cost{\B}{w} &= \costpas{\B}{w} + \costact{\B}{w}
\end{align*}

\noindent
Our notion of active cost is related to the concept of \emph{total
communication cost} applied to measure performance of distributed
algorithms.
Total communication cost measures the number of bits transferred in
order to complete a given protocol.
We simplify this notion and adapt it to the theoretical concepts as
follows.
Noting that the constructions use local asynchronous automata and
trivial message content,
as active cost, we simply count the number of participants in each
communication.
Thus we abstract away from the details of information distribution
through a network, routing information, and similar.
A more thorough evaluation of communication costs will have to consider these aspects.
For example, for asynchronous automata, (the log of) the product of the state spaces of the processes participating in a communication needs to be considered.
Without a more detailed connection to actual implementations of these abstract notions of interaction it is hard to consider the actual costs of dissemination of information.
In particular, disconnecting from a channel does not incur further cost in our model. But in a real implementation it would have to incur some communication costs. 

Our passive cost is intended to capture the attention given by the
processes to the communication.
Thus, for every process and every channel that this process is
connected to, we count a fixed cost.
This could correspond, for example, to analyzing several types of
header information on a single network to identify what to do with
messages, or connection to several networks.

\subsubsection{Costs of Fixed and Reconfigurable Automata}
Now let us use this measure to calculute the costs of the RAA $\A$
built in Section~\ref{sec:lower bound} and its fixed AA equivalent $\B$
where all processes are fully-listening.
Let $n = |\Proc|$.
We fix an ``average'' execution $w_\A$ that first sets the dependent
set from $\emptyset$ to some $D$, followed by $k$ communications on $D$.
Since there are $2^n$ possible choices for $D$, and the RAA has to
cycle through all of them in order, we choose for our average execution
the $D$ in the middle, i.e. the dependent set reached after $2^{n-1}$
communications on $\swi$.
Let $D = \{c_1, \dots, c_d\}$.
After setting the dependent set to $D$, we cycle through elements of
$D$ for  some (large) number $k$ of communications.
In the end, the execution is $w_\A = \swi^{2^{n-1}} \cdot (c_1 \dots
c_d)^{k/d}$.
On the other hand, all the machinery used to set the dependent set to
$D$ in $\A$ is not relevant for $\B$.
Therefore, in order to be fair, we omit this part of the execution when
computing the cost for $\B$, and set $w_\B = (c_1 \dots c_d)^{k/d}$.

Let us now compute $\cost{\A}{w_\A}$ and $\cost{\B}{w_\B}$.
First we start with $\B$.
Every process in $\B$ is fully-listening, so $\costpas{\B}{w_\B} = k
n^2$.
Moreover every communication involves all processes, so
$\costact{\B}{w_\B} = kn$.
Therefore the total cost for $\B$ is $\cost{\B}{w_\B} = k (n^2+n)$.

Now let us turn to $\A$.
At any moment, each process listens to either 2 or 3 channels: the
current switching channel, the channel assigned to the process, and the
channel previous to the assigned one according to the order of the
dependent set if the assigned channel belongs to it.
The exact number does not matter too much, only that it is a small
(i.e. less than $O(n)$), so for the computation of the passive cost we
just assume that each process is always connected to 3 channels.
With this assumption, we get that $\costpas{\A}{w_\A} = 3n (2^{n-1} +
k)$.
For the active cost, each communication on the switching channel in the
first half of $w_\A$ involves all processes, and all communications in
the second half only concern 2 processes, so $\costact{\A}{w_\A} =
n2^{n-1} + 2k$.
This gives a total cost of $\cost{\A}{w_\A} = k(3n+2) + 4n2^{n-1}$.

\subsubsection{Costs Analysis}
We can now compare for different values of $k$ and $n$ which model
performs better in terms of communication costs.
First, we remark that there is a large overhead in the machinery used
to create the dependent set in $\A$, which means that if $k$ is not
much larger than $n$ the fixed automaton $\B$ is cheaper, simply due to
not including the machinery in its cost.
Specifically, the cost for $\A$ starts being smaller when $k$ is in
$O(2^n/n)$, and will stay smaller for even higher values of $k$.
Moreover, when $k$ is so much larger that the second half of
$\cost{\A}{w_\A}$ is irrelevant compared to the first half,
what remains is $k(n^2+n)$ for $\B$ and $k(3n+2)$ for $\A$.
In other words, while both are understandably linear in $k$, we lower
the dependency in $n$ from squared to linear when using the
reconfigurable automaton.

In the next section, we give an alternative construction of an RAA
whose language requires full distribution that slightly increases the
passive costs in exchange for a reduced machinery cost.
We then re-analyze the cost/benefit calculation of using each of the
constructions.

\subsection{Different Trade-off between Passive and Machinery Costs}

We outline a second construction that allows for a finer control on the
set of dependent channels using an additional switching channel.
Aside from the two switching channels, there is always one
\emph{highlighted}
non-switching channel.
The first switching channel, $\toggle$ (for \emph{toggling channel}),
is used to add the highlighted channel to the dependent set.
The second switching channel, $\highl$ (for \emph{highlighting
	channel}), changes the highlighted
channel to the next non-switching channel.
With that setup, it is always possible to go from an empty dependent set
to every set $D$ in a linear number of steps by iterating on $D$:
switch the highlighted channel to the next one in $D$, then toggle it,
repeat.
This takes at most $n$ uses of $\highl$ to cycle through all channels
and $n$ uses of $\toggle$ to add them.
In the previous construction, it could take up to $2^n$ steps to reach
a target set.
Finally, after enough uses of $\highl$, the two switching channels
change so that every channel can eventually be one of the switching
channels, and the dependent set is also reset to the empty set.

Technically, this is implemented using the following notions.
\begin{itemize}
    \item 
    As before, at every given moment of time there is a set of channels on which communication is unrestricted and (the complement) set of channels on which communication happens only in a fixed (cyclic) order. 
    \item 
    As before, at every given moment of time, every process is in charge of one channel. This process ``initiates'' communication on this channel. 
    \item 
    At every given moment of time, there is one channel that is a candidate to be included in the set of restricted channels.
    Each process keeps a record of this channel in their state space.
    The coordination between all processes ensures that all processes have a common view of which channel this is at all times. 
    \item
    At every given moment of time, there are two global coordination channels: a toggling channel and a highlighting channel.
    Each process keeps a record of these channels in their state space.
    The coordination between all processes ensures that all processes have a common view of these channels.
    \begin{itemize} 
    \item
    A communication on the toggling channel adds the current highlighted channel to the set of restricted channels.
    \item 
    A communication on the highlighting channel either tells all processes that they should jointly move to highlight the next possible channel.
    Once all channels have been considered as possible channel to be added, the identity of the toggling channel and the highlighting channel is changed.
    \end{itemize}
\end{itemize}

\noindent
Formally, let us first introduce some useful notations.
We let $\Proc = \{p_1, \dots, p_n\}$ and $C = \{c_1, \dots, c_{n+2}\}$.
As in the previous construction, for every subset $D \subseteq C$ we
fix $\nextcha{}{D}: C \to C$ a function that cycles through all
elements of $D$ and is the identity on $C\setminus D$.
For convenience we write $\nextcha{d}{D}$ for $\nextcha{}{D}(d)$.
As before we also use $\prevcha{}{D}:D \to D$ the inverse function and
use the same notation.
When the two switching channels $\toggle$ and $\highl$ are known, we
denote by $C^\ast = C \setminus \{\toggle,\highl\}$ the set of
non-switching channels.
Finally, let $\minstar$ and $\maxstar$ be two channels in
$C^\ast$ such that $\minstar = \nextcha{\maxstar}{C^*}$.
Those two are used to check when we have cycled through every
non-switching channel.

We build $\A^2_p = (S_p, s^0_p, \Delta_p, L_p)$ for $p = p_k$ as
follows:
\begin{itemize}
	\item $S_p = \{(c, \toggle, \highl, h, D, d) \mid  c, \toggle,
	\highl, h \in C, D \subseteq C \setminus \{\toggle,\highl\}, d \in D
	\cup \{c\}\}$ is the set of states and $s_p^0 = (c_k, c_{n+1},
	c_{n+2}, \minstar, \emptyset, c_k)$ is the initial state.
	As in the first construction, $c$
	is the channel currently assigned to $p$, initialized to $c_k$, which
	may change when the
	switching channels are replaced. The channels $\toggle$ and $\highl$
	are those two switching channels. Channel $h$ is the highlighted
	non-switching channel. Finally, $D$ is the current dependent set and
	$d$ the next channel available for a communication for $p$.
	As in the previous construction, the behavior for all $D$ such that
	$|D|\leq 1$ is the same.
	\item All processes listen to both switching channels and their
	assigned
	channel, plus, potentially, the previous channel in $D$ if $D$
	contains
	the assigned channel:

	$$L_p(c, \toggle, \highl, h, D, d) =
	\left \{ \begin{array}{l l}
		\{\toggle, \highl,c, \prevcha{c}{D}\} & \mbox{if }c \in D\\
		\{\toggle, \highl, c\} & \mbox{if }c \notin D
	\end{array} \right.$$

	\item The transition $\Delta_p$ is the union of the following seven sets:
	\begin{align}
		&\{(c, \toggle, \highl, h, D, c), c, (c, \toggle, \highl,	h, D,
		\prevcha{c}{D})\} \\
		&\{(c, \toggle, \highl, h, D, \prevcha{c}{D}), \prevcha{c}{D},
		(c,\toggle, \highl, h, D, c) \}
	\end{align}
	The first two transitions are similar to the first two of the
	previous construction.
	\begin{align}
		&\{ (c, \toggle, \highl, h, D, d), \toggle, (c, \toggle, \highl, h,
		E, \prevcha{c}{E}) \mid c\neq h \wedge E=D\cup \{h\}\}\} \label{equation:toggle mechanism first}\\
		&\{ (c, \toggle, \highl, h, D, d), \toggle, (h, \toggle, \highl, h,
		D \cup \{h\}, h) \mid c=h\}\label{equation:toggle mechanism second}
	\end{align}
	These transitions are the toggling mechanism, which adds the
	currently highlighted channel $h$ to the dependent set.
	If the followed channel is not $h$ then update the previous channel
	according to the set $D\cup \{h\}$ (\ref{equation:toggle mechanism first}).
	If the followed channel is $h$ then expect the next communication on
	$h$ (\ref{equation:toggle mechanism second}).

	The remaining are the highlighting mechanism.
	When a communication on $\highl$ occurs, the highlighted channel
	changes to the next one.
	\begin{align}
		&\{ (c, \toggle, \highl, h, D, d), \highl, (c, \toggle, \highl,
		\nextcha{h}{C^*}, D, d) \mid h \neq \maxstar\}
	\end{align}
	If the highlighted channel is not maximal then go to the next
	highlighted channel.
	\begin{align}
		&\{(c, \toggle, \highl, \maxstar, D, d), \highl, (c, \highl,
		\nextcha{\highl}{C}, \minstar, \emptyset, c) \mid c\neq
		\nextcha{\highl}{C} \}\label{equation:change switching channels first}\\
		&\{ (c, \toggle, \highl, \maxstar, D, d), \highl, (\toggle, \highl,
		\nextcha{\highl}{C}, \minstar, \emptyset, \toggle)	\mid c =
		\nextcha{\highl}{C}\}\label{equation:change switching channels second}
	\end{align}
	If we have already cycled through all non-switching channels once (the
	current highlighted channel is maximal),
	then new channels assume the roles of
	toggling and highlighting, and the highlighted channel
	and dependent set are both reset.
	We may also have to update the first component if the channel
	assigned to
	this process becomes a switching channel.
	That is, either the assigned channel is untouched by the change (\ref{equation:change switching channels first}),
	or
	it is assigned to $\highl$ (\ref{equation:change switching channels second}).
\end{itemize}

\noindent
We give an illustration of this construction in
Figure~\ref{example:construction2}.
As in the previous example, we have two processes $p$ and $q$ whose
initial assigned channels are $1$ and $2$ respectively.
Channels $3$ and $4$ serve as the toggling and highlighting channels.
Following the upper path toggles both $1$ and $2$ and 
adds them to the dependent set, eventually reaching the blue states
where those two have to be played in the correct order.
The lower path skips toggling $1$ and therefore does not reach a state
with more than one channel in $D$.
In any case, after the highlighting channel is used twice, the two
switching channels change.
This leads to the state in red where $4$ and $1$ are the new toggling
and highlighting channels respectively.
For $p$ this means that its assigned channel becomes $3$, while $q$'s
does not change.
The path then continues, crossing into new areas of the automaton with
each switching channels change, and eventually leads back to the
initial state.

\begin{figure}
	\tikzset{every state/.style={minimum size=15pt}}
	\begin{center}
		\begin{tikzpicture}[
			auto,
			node distance=1.5cm,
			semithick
			]
			\begin{scope}
				\node[draw,rounded corners,initial left,initial text=] (p1)
				{$1, 3, 4, 1, \emptyset, 1$};
				\node[] (Lp1) [below = 0.cm of p1] {$1,3,4$};
				\node[draw,rounded corners] (p2) [above right = 0.5cm of p1]
				{$1, 3, 4, 1, \{1\}, 1$};
				\node[] (Lp2) [below = 0.cm of p2] {$1,3,4$};
				\node[draw,rounded corners] (p3) [right = 0.5cm of p2] {$1, 3,
				4, 2, \{1\}, 1$};
				\node[draw,rectangle] (p) [above = 1.2cm of p3] {$p$};
				\node[] (Lp3) [below = 0.cm of p3] {$1,3,4$};
				\node[draw,rounded corners,color=blue] (p4) [below right =
				0.5cm of p3] {$1, 3, 4, 2, \{1,2\}, 2$};
				\node[] (Lp4) [right = 0.cm of p4] {$1,2,3,4$};
				\node[draw,rounded corners,color=blue] (p5) [above = 1cm of p4]
				{$1, 3, 4, 2, \{1,2\}, 1$};
				\node[] (Lp5) [right = 0.cm of p5] {$1,2,3,4$};
				\node[draw,rounded corners,color=red] (p6) [below = 2cm of p4]
				{$3, 4, 1, 2, \emptyset, 3$};
				\node[] (Lp6) [below = 0.cm of p6] {$1,3,4$};
				\node[] (pdots1) [left of =p6] {$...$};
				\node[draw,rounded corners] (p7) [below right = 0.5cm of p1]
				{$1, 3, 4, 2, \emptyset, 1$};
				\node[] (Lp7) [below = 0.cm of p7] {$1,3,4$};
				\node[draw,rounded corners] (p8) [right = 0.5cm of p7] {$1, 3,
				4, 2, \{2\}, 1$};
				\node[] (Lp8) [below = 0.cm of p8] {$1,3,4$};

				\path[->] (p1) edge [loop above] node [above] {$1$} (p1);
				\path[->] (p1) edge [] node [above left] {$3$} (p2);
				\path[->] (p2) edge [loop above] node [above] {$1,3$} (p2);
				\path[->] (p2) edge [] node [above] {$4$} (p3);
				\path[->] (p3) edge [loop above] node [above] {$1$} (p3);
				\path[->] (p3) edge [] node [above right] {$3$} (p4);
				\path[->,color=blue] (p4) edge [bend right] node [right] {$2$}
				(p5);
				\path[->] (p4) edge [loop below] node [below] {$3$} (p4);
				\path[->,color=blue] (p5) edge [bend right] node [left] {$1,3$}
				(p4);
				\path[->,color=red] (p4) edge [bend left] node [right] {$4$}
				(p6);
				\path[->,color=red] (p5.south east) edge [bend left] node
				[right] {$4$} (p6);
				\path[->,color=red] (p3) edge [] node [above right] {$4$} (p6);
				\path[->] (p6) edge [loop right] node [right] {$3$} (p6);
				\path[->] (p6) edge (pdots1);
				\path[->] (p1) edge [] node [above right] {$4$} (p7);
				\path[->] (p7) edge [loop above] node [above] {$1$} (p7);
				\path[->] (p7) edge [] node [above] {$3$} (p8);
				\path[->] (p8) edge [loop above] node [above] {$1,3$} (p8);
				\path[->,color=red] (p7) edge [] node [below right] {$4$} (p6);
				\path[->,color=red] (p8) edge [] node [below] {$4$} (p6);

				\draw[dashed, rounded corners] (-1.2,2.3) rectangle (10.8,-1.8)
				{};
				\draw[] (1,2.5) node {$(\toggle,\highl) = (3,4)$};
				\draw[dashed, rounded corners] (5.2,-1.9) rectangle (10.8,-3.5)
				{};
				\draw[] (7,-3.7) node {$(\toggle,\highl) = (4,1)$};
				\draw[->,color=red] (5.5,-2.6) -- (4.7,-2.6);
				\draw[dashed, rounded corners] (2.1,-1.9) rectangle (5,-3.5) {};
				\draw[] (3.5,-3.7) node {$(\toggle,\highl) = (1,2)$};
				\draw[] (3.5,-2.6) node {$\dots$};
				\draw[->,color=red] (2.5,-2.6) -- (1.7,-2.6);
				\draw[dashed, rounded corners] (-1.2,-1.9) rectangle (2,-3.5)
				{};
				\draw[] (0.5,-3.7) node {$(\toggle,\highl) = (2,3)$};
				\draw[] (0.5,-2.6) node {$\dots$};
				\draw[->,color=red] (0.5,-1.7) -- (0.5,-1.2);
			\end{scope}
			\begin{scope}[shift={(0,-7.5)}]
				\node[draw,rounded corners,initial left,initial text=] (p1)
				{$2, 3, 4, 1, \emptyset, 2$};
				\node[] (Lp1) [below = 0.cm of p1] {$2,3,4$};
				\node[draw,rounded corners] (p2) [above right = 0.5cm of p1]
				{$2, 3, 4, 1, \{1\}, 2$};
				\node[] (Lp2) [below = 0.cm of p2] {$2,3,4$};
				\node[draw,rounded corners] (p3) [right = 0.5cm of p2] {$2, 3,
				4, 2, \{1\}, 2$};
				\node[draw,rectangle] (p) [above = 1.3cm of p3] {$q$};
				\node[] (Lp3) [below = 0.cm of p3] {$2,3,4$};
				\node[draw,rounded corners,color=blue] (p4) [below right =
				0.5cm of p3] {$2, 3, 4, 2, \{1,2\}, 2$};
				\node[] (Lp4) [right = 0.cm of p4] {$1,2,3,4$};
				\node[draw,rounded corners,color=blue] (p5) [above = 1cm of p4]
				{$2, 3, 4, 2, \{1,2\}, 1$};
				\node[] (Lp5) [right = 0.cm of p5] {$1,2,3,4$};
				\node[draw,rounded corners,color=red] (p6) [below = 2cm of p4]
				{$2, 4, 1, 2, \emptyset, 2$};
				\node[] (Lp6) [below = 0.cm of p6] {$2,3,4$};
				\node[] (pdots1) [left of =p6] {$...$};
				\node[draw,rounded corners] (p7) [below right = 0.5cm of p1]
				{$2, 3, 4, 2, \emptyset, 2$};
				\node[] (Lp7) [below = 0.cm of p7] {$2,3,4$};
				\node[draw,rounded corners] (p8) [right = 0.5cm of p7] {$2, 3,
				4, 2, \{2\}, 2$};
				\node[] (Lp8) [below = 0.cm of p8] {$2,3,4$};

				\path[->] (p1) edge [loop above] node [above] {$2$} (p1);
				\path[->] (p1) edge [] node [above left] {$3$} (p2);
				\path[->] (p2) edge [loop above] node [above] {$2,3$} (p2);
				\path[->] (p2) edge [] node [above] {$4$} (p3);
				\path[->] (p3) edge [loop above] node [above] {$2$} (p3);
				\path[->] (p3) edge [] node [above right] {$3$} (p4);
				\path[->,color=blue] (p4) edge [bend right] node [right] {$2$}
				(p5);
				\path[->] (p4) edge [loop below] node [below] {$3$} (p4);
				\path[->,color=blue] (p5) edge [bend right] node [left] {$1,3$}
				(p4);
				\path[->,color=red] (p4) edge [bend left] node [right] {$4$}
				(p6);
				\path[->,color=red] (p5.south east) edge [bend left] node
				[right] {$4$} (p6);
				\path[->,color=red] (p3) edge [] node [above right] {$4$} (p6);
				\path[->] (p6) edge [loop right] node [right] {$2$} (p6);
				\path[->] (p6) edge (pdots1);
				\path[->] (p1) edge [] node [above right] {$4$} (p7);
				\path[->] (p7) edge [loop above] node [above] {$2$} (p7);
				\path[->] (p7) edge [] node [above] {$3$} (p8);
				\path[->] (p8) edge [loop above] node [above] {$2,3$} (p8);
				\path[->,color=red] (p7) edge [] node [below right] {$4$} (p6);
				\path[->,color=red] (p8) edge [] node [below] {$4$} (p6);

				\draw[dashed, rounded corners] (-1.2,2.3) rectangle (10.8,-1.8)
				{};
				\draw[] (1,2.5) node {$(\toggle,\highl) = (3,4)$};
				\draw[dashed, rounded corners] (5.2,-1.9) rectangle (10.8,-3.5)
				{};
				\draw[] (7,-3.7) node {$(\toggle,\highl) = (4,1)$};
				\draw[->,color=red] (5.5,-2.6) -- (4.7,-2.6);
				\draw[dashed, rounded corners] (2.1,-1.9) rectangle (5,-3.5) {};
				\draw[] (3.5,-3.7) node {$(\toggle,\highl) = (1,2)$};
				\draw[] (3.5,-2.6) node {$\dots$};
				\draw[->,color=red] (2.5,-2.6) -- (1.7,-2.6);
				\draw[dashed, rounded corners] (-1.2,-1.9) rectangle (2,-3.5)
				{};
				\draw[] (0.5,-3.7) node {$(\toggle,\highl) = (2,3)$};
				\draw[] (0.5,-2.6) node {$\dots$};
				\draw[->,color=red] (0.5,-1.7) -- (0.5,-1.2);
			\end{scope}
		\end{tikzpicture}
		\caption{Illustration of the second construction for $n =
		2$.}\label{example:construction2}
	\end{center}
\end{figure}

\noindent
Let $\A^2$ be the parallel composition of $(\A^2_p)$.
As with the previous construction we show that for every equivalent AA,
processes are either listening to all channels or can be led to a
trivial component.

\begin{lem}\label{lemma:main2}
	Let $\B$ be an AA such that $\Lang(\B) = \Lang(\A^2)$. Each process
	in $\B$ is either fully-listening or trivializable.
\end{lem}

\begin{proof}
	Similarly to the previous construction, let $\B = ((S_p)_{p \in
	\Proc}, (s^0_p)_{p \in \Proc}, (\delta_a)_{a \in \Alp})$, and let $p
	\in \Proc$.
	Again, assume that $p$ is not fully-listening, so let $C_p =
	\dom^{-1}(p)
	\subsetneq C$ and let $c \in C \setminus C_p$ be a channel that $p$
	does
	not listen to.

	Start from some state $s_p$ reachable by $p$ in $\B$.
	By definition, there is a computation $w$ in $\Lang(\B)$ such that $p$
	reaches state $s_p$ after $w$.
	Consider the same computation $w$ in $\A^2$. The computation ends in
	some configuration where $(\toggle,\highl)$ are the current toggling
	and highlighting channels respectively.
	We want to find a continuation of this computation where $c$ is the
	highlighting channel afterwards.
	To that end, let $\highl \cdot c_1 \cdot ... \cdot c_{n-1} \cdot c$
	be the sequence of
	channels from $\highl$ to $c$ according to the order from
	$\nextcha{}{C}$.
	Then there exists a continuation $w'$ of the form $w' = \highl^{k_0}
	\cdot c_1^{k_1} \cdot ... \cdot c_{n-1}^{k_{n-1}}$ with $k_0, \dots,
	k_{n-1} \in \N$ such that:
	\begin{itemize}
		\item $ww' \in \Lang(\A)$,
		\item $c$ is the current highlighting channel, and the dependent
		set $D$ is $\emptyset$.
	\end{itemize}
	Note that we do not make assumptions regarding the toggling channel
	$\toggle$.

	We show that every possible continuation $w'' \in (C \setminus
	\{c\})^\ast$ must be possible in $p$.
	Consider such an arbitrary continuation $w'' \in (C \setminus
	\{c\})^\ast$.
	If $w''$ involves no communication on $\toggle$, then the dependent
	set
	stays empty and none of the switching channels change, so $ww'w''$
	still belongs to $\Lang(\A)$. In particular, $p$ cannot block
	communications in $C_p$.
	If $w''$ does involve communications on $\toggle$, then after the
	first
	such communication the dependent set is updated to $\{\minstar\}$.
	Afterwards, the dependent set does not change anymore and neither do
	the switching channels.
	Since the dependent set is only a singleton, all communications on
	non-switching channels are allowed to go through in every possible
	order.
	Therefore in that case as well $ww'w'' \in \Lang(A)$. Again, $p$
	cannot
	block communications in $C_p$.

	Since it holds for every such continuation, it holds \textit{a
		fortiori} for all continuations $w'' \in C_p^*$.
	Therefore $p$ is indeed in a strongly connected component
	implementing all transitions in its domain.
\end{proof}

\noindent
We note that it is possible to modify the last construction so that
only the $\highl$ channel is dynamic and the $\toggle$ channel is
fixed.
This, however, complicates the notations further.

As before, the AA $\B$ has an alternative initial configuration making
all non fully listening processes trivial.

\begin{lem}
	The language $\Lang(\A^2)$ is repetitive.
	\label{lemma:repetitive2}
\end{lem}

\begin{proof}
	The proof is similar to the proof of Lemma~\ref{lemma:repetitive}.
	By repeatedly communicating on the highlighting channel $\highl$, the
	highlighting and toggling channels change.
	Repeated communication on the next highlighting channel leads to the
	change occurring again.
	After $n+2$ changes of the toggling and the highlighting channel, the
	same channels become the highlighting and toggling channels again.
	We call the highlighting and toggling channels returning to their
	initial values \emph{a loop}.
	If during a loop a process is ``in charge'' of channel $c$, then in
	the next loop the same process is ``in charge'' of
	$\prevcha{(\prevcha{c}{C})}{C}$.
	So, if $|C|$ is even, after $|C|/2$ loops all processes are back in
	charge of their original channels.
	If $|C|$ is odd, after $|C|$ loops all processes are back in charge
	of their original channels.
	In both cases, all processes return to their original initial states.
\end{proof}

\begin{lem}
	Let $\B$ be an AA such that $\Lang(\B) = \Lang(\A^2)$.
	There exists an alternative initial configuration
	$\vec{t}=(t_0)_{p\in\Proc}$ such that
	$\Lang(\B(\vec{t}))=\Lang(\A^2)$
	and each process in $\B(\vec{t})$ is either fully-listening or
	trivial.
	\label{lemma:trivial2}
\end{lem}

\begin{proof}
	The proof is identical to the proof of Lemma~\ref{lemma:trivial}.
\end{proof}

\subsubsection*{Communication Costs}
Finally, we compute the communication cost of this new automaton $\A^2$
and compare it to the previous one.
As explained previously, we only need a linear amount of operations to
set the dependent set to any $D$, as opposed to exponential in the
previous construction.
Therefore, we set $w_2 = (\toggle \cdot \highl)^{n/2} \cdot (c_1 \dots
c_d)^{k/d}$ to represent an ``average'' execution.
With a similar computation as with the first construction, we get that
$\costpas{\A^2}{w_2} = 4n(n + k)$ and $\costact{\A^2}{w_2} = n^2 + 2k$,
for a total cost of $\cost{\A^2}{w_2} = k(4n+2) + 5n^2$.

Compared with the first construction's cost of $\cost{\A}{w_\A} =
k(3n+2) + 4n2^{n-1}$, we see a small increase in the factor of $k$, due
to the additional switching channel increasing the passive cost.
In exchange, we obtain a massive reduction in the cost of the machinery
needed to set up the dependent set, going from exponential to simply
squared.
First we compare $\A^2$ with the equivalent fixed automaton $\B$.
Remember that the cost for $\B$ was $\cost{\B}{w_\B} = k (n^2+n)$.
For very small values of $n$, meaning up to $3$, $\B$ will be more
efficient for any value of $k$.
Starting from $n = 4$, $\A^2$ starts being more efficient when $k$ is
high enough:
\begin{center}
\begin{table}[htbp]
  \centering
  \begin{tabular}{l *{14}{c}}
    \toprule
    Value of $n$ & 1 & 2 & 3 & 4 & 5 & 6 & 7 & 8 & 9~--~11 & 12~--~14 \\
    \midrule
    $\cost{\A^2}{w_2}$ is lower for $k$ at least & --- & --- & --- & 40 & 16 & 12 & 10 & 9 & 8  & 7 \\
    \bottomrule
  \end{tabular}
\end{table}    
\end{center}
Specifically, $\A^2$ starts being more efficient than $\B$ when $k \geq
5n^2/(n^2-3n-2)$, so for high values of $n$ only \textasciitilde5
communications are enough to offset the set-up cost.
Now comparing $\A$ and $\A^2$, the slightly larger factor for $k$ in
$\A^2$ makes it less efficient when $k$ is really large compared to
$n$, specifically when $k \geq 4n2^{n-1} - 5n^2$.
Still, in both cases the factor of $k$ is linear in $n$, so we argue
that the difference between the two is minimal.

\section{Other Reconfiguration Approaches}
\label{sec:other reconfiguration}

While the connection to Zielonka / Asynchronous automata has been discussed in the introduction and, in a sense, is the main topic of this paper, we would like discuss briefly reconfiguration in a more general context. 
Here we discuss other approaches to reconfiguration and the general issues that arise when considering reconfiguration in the larger sense of the word.

First, in general formal methods ``model-based approaches'', the concepts of reconfiguration and dynamism appear slightly differently than in our work. 
We include several examples below.

\smallskip
\noindent
\emph{Dynamic Timed Automata} are timed automata that change their transitions during run time. This does not affect the communication \cite{axioms12030230}.

In order to reason about them the automata are translated directly into timed automata, where model checking and other analysis can be applied. This is similar to our work in the sense that the advanced features can be compiled away. That is, in the same way we can translate RAA to AA.

In their case, the justification for using dynamism is convenience of modelling but analysis is done on the original model.
In our case, the justification for using reconfiguration is in measuring the different communication costs. But, throughout our work, we do the analysis on top of the more expressive model claiming that it does not makes sense to just reduce it to AA and the like. 

\smallskip
\noindent
\emph{Reconfigurable Hierarchical Timed Automata} are similar to dynamic timed automata but the approach is derived from software engineering \cite{8913890}. 
As before, they introduce modelling convenience through high level syntactical elements that can be compiled away into the original model. 

\smallskip
\noindent
\emph{Reconfigurable Asynchronous Logic Automata} are a variant of cellular automata \cite{10.1145/1706299.1706301}. 
The grid and communication between different cells is fixed so reconfiguration is used in a very different sense than in the current work.

\smallskip
\noindent
\emph{Dynamic Communication Automata and Dynamic Reactive Modules} include two types of automata that are allowed to create new automata.
Dynamic Communication Automata are automata that communicate over unbounded buffers \cite{10.1007/978-3-642-37064-9_17}.
All problems about the general model are undecidable and here also creation of automata is added, making the problem even more infinite. There are some complexity results on analysis in restricted cases.
As before, the notion of dynamism is quite different from that discussed in the current work. 
Dynamic Reactive Modules combine the creation of new automata with the transfer of pointers to the variables of different automata \cite{DBLP:conf/concur/FisherHNPSV11}.
As before, the issue of creation leads to a model on which reasoning is undecidable and the main purpose of this work is to show coherent definitions and opportunities.
The usage of pointers and pointer de-referencing allows, in principle, to change during runtime the connection between different parts of the system.
However, the communication is done strictly by variable sharing and the ideas of changeable communication is obfuscated by the creation of new automata. 
Reconfiguration per-se is not studied. 

\smallskip
\noindent
\emph{Reconfigurable Automata Networks} are related to interface automata \cite{DBLP:conf/sigsoft/AlfaroH01} but include also behavior of the components (cf.~\cite{DBLP:conf/calco/GianolaKS17} for examples).
The reconfiguration relates to the abilities of the formalism to reuse components and interfaces in different ways and does not support reconfiguration during runtime.

Second, looking further afield from ``model-based approaches'', we would like to recognize a relation between temporal  graphs and graph rewriting systems to our ideas of reconfiguration.

\smallskip
\noindent
\emph{Temporal Graphs} are graphs in which edges are present only at specific times, typically modelled either as time-stamped edge sets or as sequences of graph snapshots \cite{Casteigts01102012}. 
They incorporate the timing of interactions directly into the graph structure and, thus, the dynamism is exogenous.
The central focus is on efficiently computing graph properties (such as connectivity or reachability) over the evolution. Our reconfigurable systems can be seen as generative models of temporal graphs, where connectivity evolves according to local rules. Our interest is in properties that hold over all evolution rather than efficient computation over a fixed instance. 

\smallskip
\noindent
\emph{Graph Rewriting Systems} are formal models in which graphs evolve by the application of local rewriting rules that replace a subgraph matching a given pattern with another graph, sometimes under constraints \cite{Ehrig2006}.
The studies of such systems concentrate on reachability of graph configurations or termination.
Dually, our reconfiguration is more operational state-based change and the analysis targets the evolution of the entire system and interested in the computation path.

\smallskip

Our analysis of translation between models would be akin to considering a global structure in which analysis questions about the formalism can be expressed.
While our transformation shows that the translation between AA and RAA is possible and effective and that its cost is in the dissemination of information, which we try to quantify, embedding ``dynamism'' into a single structure in other contexts might involve a huge blow up of the artefact. 

\section{Conclusion and Discussion}
\label{sec:conc}

We study the addition of reconfiguration of communication to
asynchronous
automata.
We show that in terms of expressiveness, the addition does not change
the power of the model: every language recognized distributively by
automata with reconfigurable communication can be recognized
essentially by the same automata with fixed communication.
For deterministic automata this also means that the two are bisimilar.
The same is (obviously) true in the other direction.
However, the cost of conversion is in disseminating widely all the
information and leaving it up to the processes whether to use it or not.
We also show that this total dissemination cannot be avoided.
Processes who do not get access to the full information about the
computation become irrelevant and in fact do not participate in the
distributed computation.
Thus, the conversion leads to an increase in communications that can be
avoided with reconfigurability.
We compute a cost measure for communications and show when
reconfigurable automata perform better than their fixed counterpart.

The issues of mobile and reconfigurable communication raise a question
regarding ``how much'' communication is performed in a computation.
Given a language recognized by an asynchronous automaton
(distributively),
the independence relation between letters is fixed by the language.
It follows that two distributed systems in the form of asynchronous
automata accepting (distributively) the same language must have the
same independence relation between letters.
However, this does not mean that they agree on the distribution of the
alphabet.
In case of two different distributed alphabets, what makes one better
than the other?
This question becomes even more important with systems with
reconfigurable communication interfaces.
Particularly, in reconfigurable asynchronous automata, the connectivity
changes
from state to state, which makes comparison even harder.
How does one measure (and later reduce or minimize) the amount of
communication in a system while maintaining the same behavior?
To start answering these questions, we introduced a measure of cost of
communication that is partitioned between passive and active costs.
The passive cost corresponds to the machinery that is used and the
active costs corresponds to the actual interaction.
For this notion of cost, we show some tradeoffs between AA and
different RAA for the same languages.
We leave further studies of the \emph{cost of communication} as an
interesting question for further research.

The issues of ``who is connected'' and ``with whom information is
shared''
also have implications for security and privacy.
Reconfiguration allowed us to share communication only with those who
``need to know''.
Fixed topology forced us to disseminate information widely.
If we intend to use language models and models of concurrency in
applications that involve security and privacy we need a way to reason
about dissemination of information and comparing formalisms also based
on knowledge and information.

\smallskip
\noindent
{\bf Acknowledgments}

\noindent
We would like to thank Y. Abd Alrahman and L. Di Stefano for fruitful
discussions and suggestions; A. Muscholl for highlighting the
nondeterminism of RAAs;
and an anonymous reviewer for critisizing the notion of trivializable
leading to the results on full triviality.

\clearpage

\bibliographystyle{alphaurl}
\bibliography{bib.bib}

	\end{document}